\documentclass[pre,twocolumn,showpacs]{revtex4}

\usepackage{amssymb,amsmath}

\usepackage{graphicx}% Include figure files

\usepackage{color}

\newcommand{\vicente}[1]{{ #1}}
\newcommand\beq{\begin{equation}}
\newcommand\eeq{\end{equation}}
\newcommand\beqa{\begin{eqnarray}}
\newcommand\eeqa{\end{eqnarray}}
\newcommand{\dd}{\text{d}}

\newcommand{\al}{\alpha}

\begin{document}
\title{Generalized transport coefficients for inelastic Maxwell mixtures under shear flow}
\author{Vicente Garz\'{o}\footnote[1]{Electronic address: vicenteg@unex.es;
URL: http://www.unex.es/eweb/fisteor/vicente/}}
\affiliation{Departamento de F\'{\i}sica and Instituto de Computaci\'on Cient\'{\i}fica Avanzada (ICCAEx), Universidad de Extremadura, E-06071 Badajoz, Spain}
\author{Emmanuel Trizac\footnote[2]{Electronic address: trizac@lptms.u-psud.fr;
URL: http://www.lptms.u-psud.fr/membres/trizac/}} \affiliation{Laboratoire de Physique
Th\'eorique et Mod\`eles Statistiques (CNRS UMR 8626), B$\hat{a}$timent 100,
Universit\'e Paris-Sud, 91405 Orsay cedex, France}

\begin{abstract}
The Boltzmann equation framework for inelastic Maxwell models is considered to determine the transport
coefficients associated with the mass, momentum and heat fluxes of a granular binary
mixture in spatially inhomogeneous states close to the simple shear flow. The Boltzmann equation is solved by means of a Chapman-Enskog-like expansion around the
(local) shear flow distributions $f_r^{(0)}$ for each species that retain all the hydrodynamic orders in the shear rate. Due to the anisotropy induced by the shear flow, tensorial
quantities are required to describe the transport processes instead of the conventional scalar coefficients. These tensors are given in terms of the solutions of a set of
coupled equations, which can be analytically solved as functions of the
shear rate $a$, the coefficients of restitution $\alpha_{rs}$ and the parameters of the mixture (masses, diameters and composition). Since the reference distribution
functions $f_r^{(0)}$ apply for arbitrary values of the shear rate and are not restricted to weak dissipation, the corresponding generalized coefficients turn out to be nonlinear
functions of both $a$ and $\alpha_{rs}$. The dependence of the relevant elements of the three diffusion tensors on both the shear rate and dissipation is illustrated
in the tracer limit case, the results showing that the deviation of the generalized transport coefficients from their forms for vanishing shear rates is in general
significant. A comparison with the previous results obtained analytically for inelastic hard spheres by using Grad's moment method is carried out showing a good agreement over a wide range of values for the coefficients of restitution. Finally, as an application of the theoretical expressions derived here for the transport coefficients, thermal diffusion segregation of an intruder immersed in a granular gas is also studied.
\end{abstract}

\date{\today}
\maketitle

\section{Introduction}
\label{sec1}

Granular media under rapid flow conditions are amenable to a fruitful modelization through
a gas of inelastic hard spheres (IHS) \cite{BP04}. In the simplest model, the grains are assumed to be
smooth so that the inelasticity is characterized through a constant (positive) coefficient of normal restitution $\al \leq 1$ that only affects the translational degrees of
freedom of the grains. The case $\al=1$ corresponds to elastic collisions. Due to the kinetic-energy dissipation in collisions, energy must be externally injected to the granular
gas in order to maintain it in rapid flow regime (fluid-like description). In some cases, the system is driven into the flow through a (linear) shear field (simple or uniform
shear flow, USF) where a steady state is achieved when the energy dissipated by collisions is balanced by the energy supplied by shearing work. The study of the rheological
properties in the steady USF has received consequential attention in the past years \cite{C90,Go03}, especially in the case of monodisperse granular gases.

The USF state is defined by a constant density $n$, a uniform granular temperature $T$, and a linear velocity profile $u_x=ay$, where $a$ is the constant shear rate. In the
steady state, the system admits a non-Newtonian description \cite{SGD04,AS07} characterized by shear-rate dependent viscosity and normal stress differences. An interesting
problem is the analysis of momentum and heat transport in spatially inhomogeneous states close to the USF. The physical situation is such that the granular gas is in a
strongly sheared state that deviates from the USF conditions by \emph{small} spatial gradients. The response of the system to these perturbations gives rise to additional contributions
to the momentum and heat fluxes, which can be characterized by generalized shear-rate dependent transport coefficients. Due to the mathematical difficulties met in obtaining those
coefficients from the Boltzmann collision operator for IHS \cite{BP04}, the inelastic version of the BGK model \cite{BDS99} was considered to determine the above generalized transport
coefficients \cite{L06,G06}. On the other hand, explicit expressions for these coefficients were derived by assuming particular perturbations where the steady state conditions of
the USF apply \cite{L06,G06}. This allowed us to perform a linear stability analysis of the hydrodynamic equations with respect to the USF state \cite{G06} to get the conditions
for instability at long wavelengths. The results derived for IHS from the BGK model has been then revisited by considering a mean field version of the hard sphere system where
randomly chosen pairs of particles collide with a random impact direction. This assumption, which yields a Boltzmann collision operator with a collision rate \emph{independent}
of the relative velocity of the two colliding particles, opens the possibility of obtaining exact results for granular gases in the context of the Boltzmann kinetic equation.
The above interaction model is referred to as the inelastic Maxwell model (IMM) \cite{BCG00,CCG00,EB02,BK03,Ernst1,Ernst2,TK03,GS11} and it has been widely considered by
physicists and mathematicians alike in the past few years to unveil in a clean way the role of dissipation in granular flows. In particular, the use of IMM  allows us
to analytically determine the set of generalized transport coefficients around the USF state for general unsteady conditions \cite{G07}.

All the above results refer to monocomponent granular gases. However, a real granular system is generally characterized by some degree of polydispersity in density and size
(granular mixtures). \vicente{Needless to say, the difficulties for obtaining explicit expressions of the transport coefficients increase considerably when one considers multicomponent systems since not only the number of transport coefficients is larger than for a single gas but they are also functions of more parameters such as composition, masses, sizes
and different coefficients of restitution. In the case of states close to the homogeneous cooling state, explicit forms of the Navier-Stokes transport coefficients have been derived for IHS \cite{NSIHS} by considering the so-called first Sonine approximation while exact expressions of these coefficients have been also obtained for IMM \cite{GA05}. In the case of far from equilibrium states, the results for granular mixtures are more scarce. In particular, the rheological properties (shear stress and normal stress differences) of inelastic Maxwell mixtures under USF has been explicitly determined in terms of the parameters of the mixture (concentration, masses, diameters and coefficients of restitution) \cite{G03,GT10}. As in the case of monocomponent granular gases \cite{G07}, the use of the Boltzmann collision operator of IMM allows in principle to determine the transport
properties in a strongly sheared granular mixture without introducing additional and sometimes uncontrolled approximations. In addition, as has been mentioned in previous
papers, the results derived for inhomogeneous states from IMM compare well (especially in the case of low order moments) with those obtained from IHS \cite{GA05,G03,S03,MGV14}, showing the reliability of IMM to assess the impact of collisional dissipation in granular flows.}

\vicente{The goal of this paper is to study} mass, momentum and heat transport in a strongly sheared binary mixture. In this case and taking the USF state as the reference one, the set of
Boltzmann kinetic equations for the mixture is solved by means of a Chapman-Enskog-like expansion \cite{L06} around the distributions $f_r^{(0)}$ of each species. Since the above
distributions hold for arbitrary values of the shear rate \cite{GT10}, the different approximations in the Chapman-Enskog method retain all the hydrodynamic orders in $a$. Thus,
the non-equilibrium problem analyzed here accounts for two kinds of spatial gradients: \emph{small} gradients due to the (slight) perturbations to the USF and arbitrarily \emph{large}
shear rates due to the reference shear flow state. In this paper, we will restrict our calculations to first order (Navier-Stokes-like hydrodynamic order) in the spatial gradients
of concentration, temperature and flow velocity. \vicente{It is important to remark that although the form of the zeroth-order distributions $f_r^{(0)}$ is not known, we only need their second- and fourth-degree velocity moments to evaluate transport around USF. The use of IMM instead of IHS allows us to \emph{exactly} get these moments without the explicit knowledge of $f_r^{(0)}$. This is perhaps the main advantage of considering Maxwell models (both elastic and inelastic).}

%The hydrodynamic Burnett equations (second order in the spatial gradients) for a single granular gas of IMM have been recently derived \cite{KGS14}.

In the first order of the expansion, the mass flux is characterized by the second-rank tensors $D_{ij}$ (diffusion tensor), $D_{p,ij}$ (pressure diffusion tensor) and $D_{T,ij}$
(thermal diffusion tensor), the pressure tensor is defined in terms of the fourth-rank viscosity tensor $\eta_{ijk\ell}$ while the heat flux is given in terms of the second-rank
tensors $D_{ij}''$ (Dufour tensor), $L_{ij}$ (pressure energy tensor) and $\lambda_{ij}$ (thermal conductivity tensor). The set of the above generalized transport coefficients
are nonlinear functions of the shear rate, the concentration and the mechanical parameters of the mixture (masses, sizes and coefficients of restitution). The determination of \vicente{the equations defining these} transport coefficients
is perhaps \vicente{the main goal} of the present contribution.

As in previous papers pertaining to IMM \cite{GT10,MGV14,SG07}, the velocity moments of the Boltzmann collision operator are given in terms of a collision frequency $\nu_0$. This parameter
can be seen as a free parameter of the model that can be chosen to optimize the agreement with the properties of interest of the original Boltzmann equation for IHS. Thus,
in order to correctly describe the velocity dependence of the original IHS collision rate, one usually assumes that the IMM collision rate is proportional to $T^{\beta}$ with $\beta=1/2$.
Here, we take $\beta$ as a generalized exponent so that different values of $\beta$ can be used to mimic different interaction potentials. We assume that $\nu_0 \propto n T^{\beta}$, with $\beta\geq 0$.
In the case $\beta=0$, $\nu_0$ is independent of temperature (model A) while when $\beta \neq 0$, $\nu_0$ is a monotonically increasing function of temperature (model B). Model A is closer to
the original model of Maxwell gases for elastic collisions \cite{TM80,GS03} while model B with $\beta=1/2$ is closer to IHS. The possibility of having a general temperature dependence
of $\nu_0(T)$ for inelastic repulsive models has been also introduced in the granular literature\cite{Ernst1,Ernst2,Y13}. One of the main features of model A
is that the reduced shear rate $a^*=a/\nu_0$ (which is the relevant parameter measuring the departure from the homogeneous cooling state) does not change in time and so,
a non-Newtonian hydrodynamic regime (where $a^*$ and the coefficients of restitution $\al_{rs}$ are independent parameters) is achieved for long times. In this regime, the
combined effect of both control parameters on the (scaled) transport coefficients can be studied analytically for model A. This is a bonus feature of this model
that contrasts with the results derived for model B where only analytical results can be obtained in the steady state limit (namely, when viscous heating and energy lost by collisions cancel each other).

\vicente{Our results show that in general the generalized transport
coefficients associated with the mass, momentum and heat fluxes are given in terms of the solutions of a set of coupled nonlinear
differential equations. In the case of model B ($\beta\neq 0$), these equations must be numerically solved with the appropriate boundary conditions to obtain their hydrodynamic forms.
On the other hand, for model A ($\beta=0$),
they reduce to a set of coupled algebraic equations that can be analytically solved. This allows us to provide explicit expressions of these coefficients in terms of the shear rate and the
parameters of the mixture. This achievement is perhaps one of the most relevant results of the present paper since it extends to binary mixtures previous results obtained for monocomponent gases \cite{G07}.
Nevertheless, the above expressions still involve quite a tedious algebra due essentially to the intricate dependence of the velocity moments of the zeroth-order distributions $f_r^{(0)}$
on both the concentration and the (reduced) shear rate. Thus, an exploration of the full parameter space, in principle straightforward, is beyond the scope of this presentation
and a reduced problem will be addressed.
%To illustrate the dependence of the transport coefficients on the parameter space,
Indeed, the tracer limit (namely, a binary mixture where the concentration of one of the species is negligible)
is specifically considered to analyze the behavior of the diffusion coefficients.}

\vicente{Some results for  diffusion around USF have been previously reported by one of the authors of the present paper. Thus,
in the tracer limit and for steady state conditions, explicit expressions of the diffusion tensors $D_{ij}$, $D_{p,ij}$ and $D_{T,ij}$ were derived for IHS \cite{G02,G07bis} by using Grad's moment method \cite{G49}. In the case of IMM, the tracer diffusion tensor $D_{ij}$ has been also obtained \cite{G03} in the steady state where the (reduced) shear rate $a^*$ is coupled to the coefficients of restitution $\alpha_{rs}$. The results derived here for IMM extend to \emph{finite} concentration the previous attempts made for the diffusion tensors in the tracer limit. In addition, given that all the previous works \cite{G03,G02,G07bis} have been restricted to the steady state, the results reported in this paper for model A (see subsection \ref{sub5a}) offers the possibility of assessing independently the influence of both $a^*$ and $\alpha_{rs}$ on the diffusion of intruders in a sheared granular gas.}

Finally, as an interesting application of the general results, a segregation
criterion based on the thermal diffusion factor is derived in the tracer limit. This criterion shows the transition between two different regions (upwards and downwards segregation) by varying the different parameters of the system. This
study complements a previous analysis carried out for IHS \cite{GV10}. Our results show that the form of the phase diagrams of
segregation is quite similar to those obtained before for IHS.

The plan of the paper is as follows. In Sec.\ \ref{sec2}, the Boltzmann equation for inelastic Maxwell mixtures is introduced and the USF problem is defined. The Chapman-Enskog-like expansion around the USF state is described in Sec.\ \ref{sec3} while Sec.\ \ref{sec4} deals with the evaluation of the generalized transport coefficients associated with the mass, momentum and heat fluxes. The tracer limit is considered in Sec.\ \ref{sec5} to illustrate the dependence of the (scaled) transport coefficients on the reduced shear rate and the coefficients of restitution. In this limiting case, the diffusion coefficients are the relevant transport coefficients of the mixture. The dependence of the coefficients
$D_{ij}$, $D_{p,ij}$ and $D_{T,ij}$ on both $a^*$ and $\al_{rs}$ is analyzed for model A for general unsteady conditions while steady state conditions are assumed to get \vicente{the
form of $D_{ij}$ for model B ($\beta\neq 0$) in order to compare with previous results derived for IHS \cite{G07bis}. Comparison shows in general good agreement even for strong dissipation.}  Thermal diffusion segregation is studied in Sec. \ref{sec6} while a brief discussion of the results reported in this paper is provided in Sec. \ref{sec7}.

\section{Inelastic Maxwell mixtures under shear flow}
\label{sec2}

\subsection{Inelastic Maxwell mixtures}

We consider a granular binary mixture modeled as an IMM. In the absence of external forces, the set of nonlinear  Boltzmann equations for the
one-particle distribution function $f_{r}({\bf r},{\bf v},t)$ of species $r$ ($r=1,2$)
reads
\begin{equation}
\label{2.1}
\left(\frac{\partial}{\partial t}+{\bf v}\cdot \nabla \right)f_{r}({\bf r},{\bf v};t)
=\sum_{s=1}^2\;J_{rs}\left[{\bf v}|f_{r}(t),f_{s}(t)\right] \;,
\end{equation}
where the Boltzmann collision operator $J_{rs}\left[{\bf v}_{1}|f_{r},f_{s}\right]$ for
IMM describing the scattering of pairs of particles is
\beqa
J_{rs}\left[{\bf v}_{1}|f_{r},f_{s}\right] &=&\frac{\omega_{rs}}{n_s\Omega_d}
\int \dd{\bf v}_{2}\int \dd\widehat{\boldsymbol {\sigma }}\left[ \alpha_{rs}^{-1}f_{r}({\bf v}_{1}')f_{s}({\bf v}_{2}')\right.
\nonumber\\
& & \left.-f_{r}({\bf v}_{1})f_{s}({\bf v}_{2})\right]
\;.
\label{2.2}
\eeqa
Here,
\begin{equation}
\label{2.3}
n_r=\int \dd{\bf v} f_r({\bf v})
\end{equation}
is the number density of species $r$, $\omega_{rs}$ is an effective collision frequency for collisions  of type $r$-$s$,  $\Omega_d=2\pi^{d/2}/\Gamma(d/2)$
is the total solid angle in $d$ dimensions, and $\alpha_{rs}\leq 1$ refers to the
constant coefficient of restitution  for collisions between particles of species $r$
with $s$.   In addition, the primes on the velocities denote the initial values $\{{\bf
v}_{1}^{\prime}, {\bf v}_{2}^{\prime}\}$ that lead to $\{{\bf v}_{1},{\bf v}_{2}\}$
following a binary collision:
\begin{subequations}
\begin{equation}
\label{2.4}
{\bf v}_{1}^{\prime }={\bf v}_{1}-\mu_{sr}\left( 1+\alpha_{rs}
^{-1}\right)(\widehat{\boldsymbol {\sigma}}\cdot {\bf g}_{12})\widehat{\boldsymbol{\sigma}},
\end{equation}
\beq
\label{2.4.1}
{\bf v}_{2}^{\prime}={\bf v}_{2}+\mu_{rs}\left(
1+\alpha_{rs}^{-1}\right) (\widehat{\boldsymbol {\sigma}}\cdot {\bf
g}_{12})\widehat{\boldsymbol{\sigma}}\;,
\end{equation}
\end{subequations}
where ${\bf g}_{12}={\bf v}_1-{\bf v}_2$ is the relative velocity of the colliding pair,
$\widehat{\boldsymbol {\sigma}}$ is a unit vector directed along the centers of the two colliding
spheres, and $\mu_{rs}=m_r/(m_r+m_s)$ where $m_r$ is the mass of a particle of species $r$.

Apart from $n_r$, the relevant quantities in a binary mixture at a hydrodynamic level are the flow velocity $\mathbf{u}$ and the granular temperature $T$. They are defined, respectively, as
\begin{equation}
\label{u}
{\bf u}=\frac{1}{\rho}\sum_{s=1}^2\;\rho_s{\bf u}_s=\sum_{s=1}^2\int
\dd{\bf v}m_s{\bf v}f_s({\bf v}),
\end{equation}
\begin{equation}
\label{T}
nT=\sum_{s=1}^2\,n_sT_s=\sum_{s=1}^2\,\int \dd{\bf v}\frac{m_s}{d}V^2f_s({\bf v}).
\end{equation}
In Eqs.\ \eqref{u} and \eqref{T}, $\rho_r=m_r n_r$ is the mass density of species $r$, $n=n_1+n_2$ is the total number density, $\rho=\rho_1+\rho_2$ is
the total mass density, and ${\bf V}={\bf v}-{\bf u}$ is the peculiar velocity. Equations (\ref{u}) and (\ref{T}) also define the (mean) flow velocity
${\bf u}_r$ and the partial temperature $T_r$ of species $r$. The partial
temperature $T_r$ measures the mean kinetic energy of species $r$. As confirmed by computer simulations \cite{computer}, experiments \cite{exp} and kinetic theory calculations \cite{GD99}
the global granular temperature $T$ is in general different from the partial temperatures $T_r$ (non-equipartition of energy).

Furthermore, the mass flux for species $r$ is defined as
\begin{equation}
{\bf j}_{r}=m_{r}\int \dd{\bf v}\,{\bf V}\,f_{r}({\bf v}),
\label{2.7}
\end{equation}
the total pressure tensor is given by
\begin{equation}
{\sf P}=\sum_{s=1}^2\,\int \dd{\bf v}\,m_{s}{\bf V}{\bf V}\,f_{s}({\bf  v}), \label{2.8}
\end{equation}
and the total heat flux is
\begin{equation}
{\bf q}=\sum_{s=1}^2\,\int \dd{\bf v}\,\frac{1}{2}m_{s}V^{2}{\bf V} \,f_{s}({\bf v}).
\label{2.9}
\end{equation}
In addition, the rate of energy dissipated due to collisions among all the species defines the cooling rate $\zeta$ as
\begin{equation}
\label{2.10}
\zeta=-\frac{1}{d n T}\sum_{r,s}\int \dd{\bf v}\;m_{r}V^{2}J_{rs}
[{\bf v}|f_{r},f_{s}]\;.
\end{equation}
At a kinetic level, it is also convenient to introduce the partial cooling rates $\zeta_{r}$, measuring the rate of energy lost by species $r$. They are defined as
\beq
\label{2.10.1}
\zeta_r=\sum_{s}\, \zeta_{rs}=-\frac{1}{dn_rT_r}\sum_s\; \int \dd \mathbf{v}\; m_r V^2 J_{rs}[f_r,f_s],
\eeq
where the second identity defines the quantities $\zeta_{rs}$. The total cooling rate $\zeta$ is given by
\beq
\label{2.10.2}
\zeta=\sum_{s=1}^2\, x_s \gamma_s \zeta_s,
\eeq
where $x_r\equiv n_r/n$ is the concentration (or mole fraction) of species $r$ and $\gamma_r\equiv T_r/T$.

\vicente{As said in the Introduction}, one of the main advantages of considering IMM is that the moments of the Boltzmann collision operator $J_{rs}[f_r,f_s]$ defined by
Eq.\ \eqref{2.2} can be exactly evaluated in terms of the distributions $f_r$ and $f_s$ without the explicit knowledge of both distributions \cite{TM80}. This property
has been exploited to determine the second-, third- and fourth-degree collisional moments for a monodisperse granular gas \cite{GS07}. In the case of mixtures, only the first-,
second-degree and third-degree collisional moments \cite{GA05} have been obtained. Their explicit forms can be found in the above papers.

The results obtained before apply regardless the specific form of the effective collision frequencies $\omega_{rs}$. These frequencies are independent of velocity but
depend on space an time through its dependence on density and temperature. On physical grounds, $\omega_{rs} \propto n_s$. As in previous works on IMM \cite{G07,SG07,GT10},
we will assume that
$\omega_{rs}\propto n_s T^{\beta}$, with $\beta\geq 0$. The case $\beta=0$ (a collision frequency
independent of temperature) will be referred as model A while the case $\beta\neq 0$ will be called model B. The
collision frequencies $\omega_{rs}$ can be seen as free parameters in the model to
optimize the agreement with some property of interest of IHS. Here, $\omega_{rs}$ is chosen to get the same partial cooling rate $\zeta_{rs}$ as for IHS (evaluated \vicente{by using a Gaussian distribution for $f_r$}). With this choice, $\omega_{rs}$ can be written as
\cite{G03,GA05}
\begin{equation}
\label{2.5}
\omega_{rs}=x_s\left(\frac{\sigma_{rs}}{\sigma_{12}}\right)^{d-1}
\left(\frac{\theta_r+\theta_s}{\theta_r\theta_s}\right)^{1/2}\nu_0, \quad \nu_0=A(\beta) n T^\beta,
\end{equation}
where the value of the quantity $A$ will be defined later (see subsection \ref{sec5b}). In Eq.\ (\ref{2.5}), $\sigma_{rs}=(\sigma_r+\sigma_s)/2$, and
\begin{equation}
\label{2.6}
\theta_r=\frac{m_r}{\gamma_r}\sum_{s=1}^2m_s^{-1}.
\end{equation}

\subsection{Uniform shear flow}

We assume that the mixture is under USF. This state is macroscopically characterized by constant partial densities, a uniform temperature,
and a linear velocity profile
\begin{equation}
\label{2.11}
{\bf u}(y)={\bf u}_1(y)={\bf u}_2(y)=ay \widehat{{\bold x}},
\end{equation}
where $a$ is the \emph{constant} shear rate. This linear velocity profile, \vicente{in computer simulations, can conveniently be generated by the Lees-Edwards boundary conditions
\cite{LE72}}, which are simply periodic boundary conditions in the local Lagrange frame
moving with the flow velocity \cite{DSBR86}. Since $n_r$ and $T$ are here uniform, then the
mass and heat fluxes vanish and the transport of momentum (measured by the pressure
tensor) is the relevant phenomenon. At a microscopic level, the USF is characterized by a velocity distribution function that becomes \emph{uniform} in the local Lagrangian
frame moving with the flow velocity $\mathbf{u}$, i.e., $f_{s}({\bf r},{\bf v},t)=f_{s}({\bf V},t)$. In that case, Eq.\
(\ref{2.1}) becomes \cite{GS03}
\begin{equation}
\label{2.12}
\frac{\partial f_1}{\partial t}-aV_y\frac{\partial f_1}{\partial V_x}=J_{11}[f_1,f_1]+J_{12}[f_1,f_2]
\end{equation}
and a similar relation for $f_2$. The relevant balance equation in the USF state is the
balance equation for the temperature. It can be obtained from Eq.\ (\ref{2.12}) and its
counterpart for species $2$; it is given by
\begin{equation}
\label{2.13}
\nu_0^{-1}\frac{\partial}{\partial t}\ln T=-\zeta^*-\frac{2a^*}{d} P_{xy}^*,
\end{equation}
where $\zeta^*\equiv \zeta/\nu_0$, $a^*\equiv a/\nu_0$, $P_{xy}^*\equiv P_{xy}/p$, $p=nT$ being the
hydrostatic pressure. Equation (\ref{2.11}) shows that the temperature changes in time
due to the competition of two opposite mechanisms: viscous heating
(shearing work) and energy dissipation in collisions.
It is apparent that, except for model A ($\beta=0$), the collision frequency
$\nu_0(T)\propto T^{\beta}$ is an increasing function of temperature, and so $a^*(t)\propto
T(t)^{-\beta}$ is a function of time. Consequently, the (reduced) pressure tensor
$P_{xy}^*$ depends on time in the hydrodynamic regime only through its dependence on
$a^*(t)$ \cite{GS03}. Therefore, for $\beta\neq 0$, after a transient regime a steady state is achieved in the long time limit when both viscous heating and collisional cooling
cancel each other and the mixture autonomously seeks the temperature at which the above balance occurs. In this steady state, the reduced shear rate and the coefficients of
restitution are not independent parameters since they are related through the steady
state condition
\begin{equation}
\label{2.12.1} a^*P_{xy}^*=-\frac{d}{2}\zeta^*.
\end{equation}
On the other hand, when $\beta=0$, $\partial_t a^*=0$ so that the reduced shear rate
remains in its initial value regardless of the values of the coefficients of
restitution $\alpha_{rs}$. As a consequence, there is no steady state (unless $a^*$
takes the specific value given by the condition (\ref{2.12.1})) and $a^*$ and
$\alpha_{rs}$ are \emph{independent} parameters in the USF problem. Moreover, it
must be also noted that the results obtained in the steady simple shear flow state are
\emph{universal} in the sense that they apply both for model A and model B, regardless
of the specific dependence of $\nu_0$ on $T$. The rheological properties for a granular
binary mixture of IMM in the steady state were obtained in Ref.\ \cite{G03}, while a
more detailed study on the rheological properties has been carried
out in Ref.\ \cite{GT10}. \vicente{In particular, the shear stress $P_{xy}^*$ can be written as
\beq
\label{2.12.2}
P_{xy}^*=-\eta^* a^*,
\eeq
where $\eta^*$ is the (scaled) \emph{nonlinear} shear viscosity of the granular mixture. The dependence of $\eta^*$ on both $a^*$ and $\alpha_{rs}$ has been
thoroughly analyzed in Ref.\ \cite{GT10} for different systems (see for instance, Figs.\ 3, 4, 5 and 7
of \cite{GT10}).   }

\vicente{Apart from the rheological properties, an interesting quantity is the temperature ratio $\gamma\equiv T_1/T_2$, which quantifies the lack of equipartition of the kinetic energy.
Obviously, $\gamma=1$ for any value of the shear rate and/or the coefficients of restitution in the case of mechanically equivalent particles ($m_1=m_2$, $\sigma_1=\sigma_2$,
and $\al_{11}=\al_{22}=\al_{12}$). Beyond this limiting case, the temperature ratio clearly differs from 1. For model A, $\gamma$ is determined from the condition
\beq
\label{2.12.3}
\gamma=\frac{x_2}{x_1}\frac{P_{1,xx}^*+(d-1)P_{1,yy}^*}{P_{2,xx}^*+(d-1)P_{2,yy}^*},
\eeq
where $P_{r,ij}^*\equiv P_{r,ij}/p$ and
\beq
\label{2.12.4}
P_{r,ij}=\int\; \dd \textbf{V}\, m_r V_i V_j f_r(\textbf{V}).
\eeq
The expressions of the partial pressure tensors ${\sf P}_r$ have been obtained analytically for model A in Ref.\ \cite{GT10}. For model B, the forms of ${\sf P}_r$ must be determined after solving numerically a nonlinear set of differential equations. On the other hand, analytical results for ${\sf P}_r$ can be obtained in the case of model B in the steady state (where models A and B yield the same results). In this situation, the temperature ratio is obtained by solving the equation \cite{GT10}
\beq
\label{2.12.5}
\gamma=\frac{x_2\zeta_2^*P_{1,xy}^*}{x_1\zeta_1^*P_{2,xy}^*},
\eeq
where $\zeta_r^*=\sum_s\; \zeta_{rs}^*$ and
\beq
\label{2.12.6}
\zeta_{rs}^*=\frac{2\omega_{rs}^*}{d} \mu_{sr}(1+\al_{rs})\left[1-\frac{\mu_{sr}}{2}(1+\al_{rs}) \frac{\theta_r+\theta_s}{\theta_r}\right].
\eeq
Here, $\omega_{rs}^*\equiv \omega_{rs}/\nu_0$. When the expressions of $\zeta_r^*$ and ${\sf P}_r^*$ are substituted into Eq.\ \eqref{2.12.5}, one gets a closed nonlinear equation for $\gamma$ whose numerical solution provides the dependence of the temperature ratio on the parameters of the problem. As expected, the extent of equipartition violation is greater when the mass disparity is large. Moreover, the predictions of IMM for $\gamma$ compare very well (see for instance, Figs.\ 2 and 3 of Ref.\ \cite{G03}) with Monte Carlo simulations for IHS \cite{MG02} for conditions of practical interest. This excellent agreement shows again the reliability of IMM to capture the main trends observed in sheared granular flows.}

\section{Chapman-Enskog-like expansion around USF}
\label{sec3}

Let us now perturb the USF by small spatial gradients. The response of
the system to those perturbations gives rise to contributions to the mass, momentum and
heat fluxes that can be characterized by generalized transport coefficients. Our
objective is to determine the shear-rate dependence of these coefficients for inelastic
Maxwell mixtures.

In order to analyze this problem we have to start from the set of Boltzmann equations
(\ref{2.1}) with a general time and space dependence. Let ${\bf u}_{0}={\sf a}\cdot
{\bf r}$ be the flow velocity of the {\em undisturbed} USF state, where the elements of
the tensor ${\sf a}$ are $a_{ij}=a\delta_{ix}\delta_{jy}$. As expected \cite{G07,GS03},
in the {\em disturbed} state the true velocity ${\bf u}$ is in general
different from ${\bf u}_0$, and hence ${\bf u}={\bf u}_0+\delta {\bf u}$, $\delta {\bf u}$
being a small perturbation to ${\bf u}_0$. As a consequence, the true peculiar velocity
is now ${\bf c}\equiv {\bf v}-{\bf u}={\bf V}-\delta{\bf u}$, where ${\bf V}={\bf
v}-{\bf u}_0$. In the Lagrangian frame moving with ${\bf u}_0$, the Boltzmann equations
(\ref{2.1}) can be written as
\begin{subequations}
\begin{equation}
\label{3.1} \frac{\partial f_1}{\partial t}-aV_y\frac{\partial f_1}{\partial
V_x}+\left({\bf V}+{\bf u}_0\right)\cdot \nabla f_1=J_{11}[f_1,f_1]+J_{12}[f_1,f_2],
\end{equation}
\begin{equation}
\label{3.2} \frac{\partial f_2}{\partial t}-aV_y\frac{\partial f_2}{\partial
V_x}+\left({\bf V}+{\bf u}_0\right)\cdot \nabla f_2
=J_{22}[f_2,f_2]+J_{21}[f_2,f_1],
\end{equation}
\end{subequations}
where here the derivative $\nabla f_r$ is taken at constant ${\bf V}$. The macroscopic
balance equations for the densities of mass, momentum and energy associated with this disturbed USF state are obtained from Eqs.\ (\ref{3.1}) and (\ref{3.2}) with the result
\begin{equation}
\label{3.3}
\partial_tn_r+{\bf u}_0\cdot \nabla n_r+\nabla \cdot (n_r\delta {\bf u})=-
\frac{\nabla \cdot {\bf j}_r}{m_r},
\end{equation}
\begin{equation}
\label{3.4}
\partial_t\delta u_i+a_{ij}\delta u_j+({\bf u}_0+\delta {\bf u})\cdot \nabla \delta u_i=-
\rho^{-1}\nabla_j P_{ij},
\end{equation}
\beqa
\label{3.5} & & \frac{d}{2}n\partial_tT+\frac{d}{2}n({\bf u}_0+\delta {\bf u})\cdot \nabla
T=-aP_{xy}\nonumber\\
& & -\frac{d}{2}T\sum_{s=1}^2\frac{\nabla \cdot {\bf j}_s}{m_s}-
\left(\nabla \cdot {\bf q}+{\sf P}:\nabla
\delta {\bf u}+\frac{d}{2}p\zeta\right),\nonumber\\
\eeqa
where the mass flux ${\bf j}_r$, the pressure tensor ${\sf P}$, the heat flux ${\bf
q}$, and the cooling rate  are defined by Eqs.\ (\ref{2.7}), (\ref{2.8}), (\ref{2.9}),
and (\ref{2.10}), respectively, with the replacement ${\bf V}\rightarrow {\bf c}$.

We assume that the deviations from the USF state are small. This means that the spatial gradients of the hydrodynamic fields are small. For elastic gases, the specific set of
gradients contributing to each flux is restricted by fluid symmetry, Onsager relations, and the form of entropy production \cite{GM84}. However, for granular gases, only fluid
symmetry applies and so there is more flexibility in the representation of the heat and mass fluxes since they can be defined in a variety of equivalent ways depending on the
choice of hydrodynamic gradients used. Some care is thus required in comparing
transport coefficients in different representations using different independent
gradients for the driving forces. Here, the concentration $x_1$, the pressure $p$, the temperature $T$, and the local flow velocity
$\delta {\bf u}$ are chosen as hydrodynamic fields.

Since the system is strongly sheared, a solution to the set of Boltzmann equations
(\ref{3.1}) and (\ref{3.2}) can be obtained by means of a generalization of the
conventional Chapman-Enskog method \cite{CC70} in which the velocity distribution
function of each species is expanded around the {\em local} version of the shear flow distribution (reference state).
This type of Chapman-Enskog-like expansion has been already considered in the case of \emph{monocomponent} granular gases to get the set of shear-rate dependent transport
coefficients of IHS \cite{L06,G06} and IMM \cite{G07}. More technical details on this method can be found in the above references.

In the context of the Chapman--Enskog method \cite{CC70}, we look for a {\em normal} solution of the form
\begin{equation}
\label{3.6} f_s({\bf r}, {\bf V},t)\equiv f_s[A({\bf r}, t), {\bf V}],
\end{equation}
where
\begin{equation}
\label{3.7} A({\bf r},t)\equiv \{x_1({\bf r},t), p({\bf r}, t), T({\bf r}, t), \delta {\bf u}({\bf r},t)\}.
\end{equation}
This special solution expresses the fact that the space dependence of the reference shear flow is completely absorbed in the relative velocity ${\bf V}$ and all other space
and time dependence occurs entirely through a {\em functional} dependence on the fields $A({\bf r}, t)$. The functional dependence (\ref{3.6}) can be made
local by an expansion of the distribution functions $f_s$  in powers of the hydrodynamic gradients:
\begin{equation}
\label{3.8} f_s[A({\bf r}, t, {\bf V}] =f_s^{(0)}({\bf V})+ f_s^{(1)}({\bf V})+\cdots,
\end{equation}
where the reference zeroth-order distribution function corresponds to the USF distribution function but taking
into account the local dependence of the concentration, pressure and temperature and the change ${\bf V}\rightarrow {\bf V}-\delta{\bf u}({\bf r}, t)={\bf c}$. The successive
approximations $f_s^{(k)}$ are of order
$k$ in the gradients of $x_1$, $p$, $T$, and $\delta {\bf u}$ but retain all the orders in the shear rate $a$.
Here, only the first-order approximation will be analyzed.

When the expansion (\ref{3.8}) is substituted into the definitions (\ref{2.7})--\eqref{2.10}, one gets the corresponding expansions for the fluxes and the cooling rate:
\begin{subequations}
\begin{equation}
\label{3.9} {\bf j}_s={\bf j}_s^{(0)}+{\bf j}_s^{(1)}+\cdots, \quad {\sf P}={\sf P}^{(0)}+{\sf P}^{(1)}+\cdots,
\end{equation}
\begin{equation}
\label{3.10} {\bf q}={\bf q}^{(0)}+{\bf q}^{(1)}+\cdots, \quad \zeta=\zeta^{(0)}+\zeta^{(1)}+\cdots.
\end{equation}
\end{subequations}
Finally, as in the usual Chapman-Enskog method, the time derivative is also expanded as
\begin{equation}
\label{3.11}
\partial_t=\partial_t^{(0)}+\partial_t^{(1)}+\partial_t^{(2)}+\cdots,
\end{equation}
where the action of each operator $\partial_t^{(k)}$ is obtained from the hydrodynamic equations
(\ref{3.3})--(\ref{3.5}). These results provide the basis for generating the Chapman-Enskog solution to the
Boltzmann equations (\ref{3.1}) and (\ref{3.2}).

\subsection{Zeroth-order approximation}

Substituting the expansions (\ref{3.8})--(\ref{3.11}) into Eq.\ (\ref{3.1}), the kinetic equation for $f_1^{(0)}$ is given by
\begin{equation}
\label{3.12} \partial_t^{(0)} f_1^{(0)}-aV_y\frac{\partial f_1^{(0)}}{\partial V_x}=J_{11}[f_1^{(0)},f_1^{(0)}]+J_{12}[f_1^{(0)},f_2^{(0)}].
\end{equation}
To lowest order in the expansion the conservation laws yield
\begin{equation}
\label{3.13}
\partial_t^{(0)}x_1=0,\quad
T^{-1}\partial_t^{(0)}T=p^{-1}\partial_t^{(0)}p= -\frac{2}{dp}a P_{xy}^{(0)}-\zeta^{(0)},
\end{equation}
\begin{equation}
\label{3.14}
\partial_t^{(0)}\delta u_i+a_{ij} \delta u_j=0.
\end{equation}
Since $f_1^{(0)}$ is a normal solution, the time derivative in Eq.\ (\ref{3.12}) can be represented more
usefully as
\begin{eqnarray}
\label{3.15}
\partial_t^{(0)}f_1^{(0)}&=&\frac{\partial f_1^{(0)}}{\partial
x_1}\partial_t^{(0)} x_1+\frac{\partial f_1^{(0)}}{\partial p}\partial_t^{(0)} p+\frac{\partial
f_1^{(0)}}{\partial T}\partial_t^{(0)} T
\nonumber\\
& & +\frac{\partial f_1^{(0)}}{\partial \delta
u_i}\partial_t^{(0)} \delta u_i\nonumber\\
&=&-\left(\frac{2}{d p}a P_{xy}^{(0)}+T
\zeta^{(0)}\right)\left(p\frac{\partial f_1^{(0)}}{\partial p}+T\frac{\partial f_1^{(0)}}{\partial
T}\right)\nonumber\\
& & +a \; \delta u_y \frac{\partial f_1^{(0)}}{\partial c_x},
\end{eqnarray}
where in the last step we have taken into account that $f_1^{(0)}$ depends on $\delta {\bf u}$ only through the
peculiar velocity ${\bf c}$. Substituting Eq.\ (\ref{3.15}) into Eq.\ (\ref{3.12}) yields the following kinetic
equation for $f_1^{(0)}$:
\beqa
\label{3.16}
& & -\left(\frac{2}{d p}a P_{xy}^{(0)}+T \zeta^{(0)}\right)\left(p\frac{\partial f_1^{(0)}}{\partial
p}+T\frac{\partial f_1^{(0)}}{\partial T}\right) -ac_y\frac{\partial f_1^{(0)}}{\partial c_x}\nonumber\\
& &
=J_{11}[f_1^{(0)},f_1^{(0)}]+J_{12}[f_1^{(0)},f_2^{(0)}].
\eeqa
A similar equation holds for $f_2^{(0)}$. Note that Eq.\ \eqref{3.16} and its corresponding counterpart for $f_2^{(0)}$ apply for both models A and B. The zeroth-order solution leads to ${\bf j}_r^{(0)}={\bf q}^{(0)}={\bf 0}$. Therefore, the most relevant velocity moments of the distributions $f_r^{(0)}$ are the partial pressure tensors ${\sf P}_1^{(0)}$ and
${\sf P}_2^{(0)}$ (defined by Eq.\ \eqref{2.12.4} by replacing $f_r \to f_r^{(0)}$). They can be obtained from Eq.\ (\ref{3.16}) and its counterpart for $f_2^{(0)}$ when one multiplies both equations by $m_r {\bf c}{\bf c}$
and integrates
over ${\bf c}$. The set of coupled differential equations defining the above partial pressure tensors have been derived in Ref.\ \cite{GT10} (see Eqs.\ (26)--(30) in this article). These equations can be explicitly solved for general unsteady conditions in the case of model A ($\beta=0$) where the tensors ${\sf P}_1^{(0)}$ and ${\sf P}_2^{(0)}$ can be expressed in terms of the parameter space of the problem (shear rate, coefficients of restitution, masses, diameters and composition). In the case of model B ($\beta \neq 0$), analytic results can be only obtained in the steady state (where we recall that the expressions of the partial pressure tensors are the same for both models). Beyond the steady state conditions, in order to have ${\sf P}_1^{(0)}$ and ${\sf P}_2^{(0)}$ for model B one has to solve numerically the set of differential equations obeying these partial pressure tensors. This task is beyond the objective of the present paper since we are mainly interested here in providing analytic results.

\vicente{As mentioned in the Introduction, the solution to Eq.\ \eqref{3.16} has not been obtained so far (even for model A where the collision frequencies $\omega_{rs}$ are independent of
the granular temperature) and hence, the precise form of the zeroth-order distribution $f_r^{(0)}$ is not known. However, an indirect information on the behavior of $f_r^{(0)}$ is given
through its velocity moments. As a matter of fact, only the second- and fourth-degree velocity moments of $f_r^{(0)}$ are required to determine the generalized transport coefficients associated
with the first-order solution $f_r^{(1)}$. As we will show in Sec.\ \ref{sec4}, while the partial pressure tensors ${\sf P}_1^{(0)}$ and ${\sf P}_2^{(0)}$ are involved in the evaluation
of the diffusion coefficients $D_{ij}$, $D_{p,ij}$ and $D_{T, ij}$ and the viscosity tensor $\eta_{ijk\ell}$, the fourth-degree velocity moments $N_{r,ijk\ell}^{(0)}$ of $f_r^{(0)}$
(defined by Eq.\ \eqref{b8}) are needed to get the coefficients $D_{ij}''$, $L_{ij}$ and $\lambda_{ij}$ associated with the heat flux. Although the forms of ${\sf P}_r^{(0)}$
($r=1,2$) are explicitly known for granular binary mixtures \cite{GT10}, the moments $N_{r,ijk\ell}^{(0)}$ are only known for the special case of monodisperse granular gases \cite{SG07}.
Given the difficulty of obtaining those moments from the true Boltzmann collision operator, one could consider a BGK-like kinetic model for granular mixtures \cite{VGS07} to evaluate them.}

\subsection{First-order approximation}

The analysis to first order in the gradients is worked out in Appendix \ref{appA}. Only
the final results are presented here. The distribution function $f_1^{(1)}$ is of the
form
\begin{equation}
\label{3.17} f_1^{(1)}={\boldsymbol {\cal A}}_{1}\cdot \nabla x_1+ {\boldsymbol {\cal B}}_{1}\cdot \nabla
p+{\boldsymbol {\cal C}}_{1}\cdot \nabla T+{\sf {\cal D}}_{1}:\nabla \delta {\bf u},
\end{equation}
where the vectors $\{{\boldsymbol {\cal A}}_{1}, {\boldsymbol {\cal B}}_{1}, {\boldsymbol {\cal C}}_{1}\}$, and
the tensor ${\sf {\cal D}}_{1}$ are functions of the true peculiar velocity ${\bf c}$. They are the solutions of
the following set of linear integral equations:
\begin{widetext}
\begin{eqnarray}
\label{3.18} -\left(\frac{2}{d p}a
P_{xy}^{(0)}+\zeta^{(0)}\right)\left(p\partial_p+T\partial_T\right){\boldsymbol {\cal A}}_{1}-& & a
c_y\frac{\partial {\boldsymbol {\cal A}}_{1}}{\partial c_x}+{\cal L}_1 {\boldsymbol {\cal A}}_{1}+
{\cal M}_1 {\boldsymbol {\cal A}}_{2}={\bf A}_{1}\nonumber\\
& & +\left(\frac{2a}{d}\frac{\partial P_{xy}^{(0)}}{\partial x_1}+p \frac{\partial
\zeta^{(0)}}{\partial x_1}\right) {\boldsymbol {\cal B}}_{1} +\left(\frac{2aT}{d
p}\frac{\partial P_{xy}^{(0)}}{\partial x_1}+T \frac{\partial \zeta^{(0)}}{\partial
x_1}\right){\boldsymbol {\cal C}}_{1},
\end{eqnarray}
\begin{eqnarray}
\label{3.19} -\left(\frac{2}{d p}a
P_{xy}^{(0)}+\zeta^{(0)}\right)\left(p\partial_p+T\partial_T\right){\boldsymbol {\cal
B}}_{1}-& &\left(\frac{2a}{d}\frac{\partial P_{xy}^{(0)}}{\partial p}+\zeta^{(0)}+p
\frac{\partial \zeta^{(0)}}{\partial p}\right){\boldsymbol {\cal B}}_{1}
- a
c_y\frac{\partial {\boldsymbol {\cal B}}_{1}}{\partial c_x}+{\cal L}_1{\boldsymbol {\cal B}}_{1}+
{\cal M}_1 {\boldsymbol {\cal B}}_{2}\nonumber\\
& &={\bf B}_{1} -\left[\frac{2aT}{d p^2}\left(P_{xy}^{(0)}-p\frac{\partial P_{xy}^{(0)}}{\partial p} \right)-
\frac{T}{p}\frac{\partial \zeta^{(0)}}{\partial p}\right] {\boldsymbol {\cal C}}_{1},
\end{eqnarray}
\begin{eqnarray}
\label{3.20} -\left(\frac{2}{d p}a
P_{xy}^{(0)}+\zeta^{(0)}\right)\left(p\partial_p+T\partial_T\right){\boldsymbol {\cal
C}}_{1}-& & \left(\frac{2a}{d
p}P_{xy}^{(0)}+\zeta^{(0)}+\frac{2a T}{d p}\frac{\partial P_{xy}^{(0)}}{\partial T}+T
\frac{\partial \zeta^{(0)}}{\partial T}\right){\boldsymbol {\cal C}}_{1}
- a c_y\frac{\partial {\boldsymbol {\cal C}}_{1}}{\partial c_x}\nonumber\\
& & +{\cal L}_1 {\boldsymbol {\cal C}}_{1}+
{\cal M}_1 {\boldsymbol {\cal C}}_{2}={\bf C}_{1}+\left(\frac{2a}{d}\frac{\partial P_{xy}^{(0)}}{\partial T}+
p\frac{\partial \zeta^{(0)}}{\partial T}\right)
{\boldsymbol {\cal B}}_{1},
\end{eqnarray}
\begin{equation}
\label{3.21} -\left(\frac{2}{d p}a
P_{xy}^{(0)}+\zeta^{(0)}\right)\left(p\partial_p+T\partial_T\right){\cal D}_{1,k\ell}-
a c_y\frac{\partial {\cal D}_{1,k\ell}}{\partial c_x}-a\delta_{ky}{\cal D}_{1,x\ell}
+{\cal L}_1 {\cal D}_{1,k\ell}+
{\cal M}_1 {\cal D}_{2,k\ell}=D_{1,k\ell}.
\end{equation}
\end{widetext}
Here, ${\bf A}_1$, ${\bf B}_1$, ${\bf C}_1$ and $D_{1,k\ell}$ are defined by Eqs.\
(\ref{a8})--(\ref{a11}), respectively. Moreover, ${\cal L}_1$ and ${\cal M}_1$ are the
linearized Boltzmann collision operators around the reference USF state:
\begin{subequations}
\begin{equation}
\label{3.22} {\cal L}_1X=-\left(J_{11}[f_1^{(0)},X]+J_{11}[X,f_1^{(0)}]+J_{12}[X,f_2^{(0)}]\right),
\end{equation}
\begin{equation}
\label{3.23} {\cal M}_1X=-J_{12}[f_2^{(0)},X].
\end{equation}
\end{subequations}
A similar equation for $f_2^{(1)}$ applies by setting $1\leftrightarrow 2$. It is
important to note that for $\beta=\frac{1}{2}$, Eqs.\ (\ref{3.18})--(\ref{3.21}) are expected to have the
same structure as that of the Boltzmann equation for IHS, except for the explicit form
of the operators ${\cal L}_s$ and ${\cal M}_s$.

Once the form of the distributions $f_r^{(1)}$ is known, the first-order corrections to
the mass flux $j_{1,i}^{(1)}$, the pressure tensor $P_{ij}^{(1)}$ and the heat flux $q_{i}^{(1)}$ can be obtained. They are given by
\begin{equation}
\label{3.24}
j_{1,i}^{(1)}=-\frac{m_1m_2n}{\rho}D_{ij}\frac{\partial x_1}{\partial r_j}-
\frac{\rho}{p}D_{p,ij}\frac{\partial p}{\partial r_j}-\frac{\rho}{T}D_{T,ij}\frac{\partial T}{\partial r_j},
\end{equation}
\begin{equation}
\label{3.25}
P_{ij}^{(1)}=-\eta_{ijk\ell} \frac{\partial \delta u_\ell}{\partial r_k},
\end{equation}
\begin{equation}
\label{3.26}
q_{i}^{(1)}=-T^2 D_{ij}''\frac{\partial x_1}{\partial r_j}-L_{ij}\frac{\partial p}{\partial r_j}
-\lambda_{ij}\frac{\partial T}{\partial r_j},
\end{equation}
where
\begin{equation}
\label{3.27} D_{ij}=-\frac{\rho}{n m_2}\int \dd{\bf c}\,c_i\;{\cal A}_{1,j}({\bf c}),
\end{equation}
\begin{equation}
\label{3.28} D_{p,ij}=-\frac{pm_1}{\rho}\int \dd{\bf c}\,c_i\;{\cal B}_{1,j}({\bf c}),
\end{equation}
\begin{equation}
\label{3.29} D_{T,ij}=-\frac{Tm_1}{\rho}\int \dd{\bf c}\,c_i\;{\cal C}_{1,j}({\bf c}),
\end{equation}
\begin{equation}
\label{3.30}
\eta_{ijk\ell}=\sum_{s=1}^2\; \eta_{s,ijk\ell},\quad \eta_{s,ijk\ell}=-m_s\int \dd{\bf c}\;
c_ic_j {\cal D}_{s,k\ell}({\bf c}),
\end{equation}
\begin{equation}
\label{3.31} D_{ij}''=\sum_{s=1}^2\; D''_{s,ij},\quad
D''_{s,ij}=-\frac{m_s}{2T^2}\int\; \dd{\bf c}\; c^2c_i {\cal A}_{s,j}({\bf c}),
\end{equation}
\begin{equation}
\label{3.32} L_{ij}=\sum_{s=1}^2\; L_{s,ij},\quad L_{s,ij}=-\frac{m_s}{2}\int\; \dd{\bf c}\; c^2c_i
{\cal B}_{s,j}({\bf c}),
\end{equation}
\begin{equation}
\label{3.33} \lambda_{ij}=\sum_{s=1}^2\; \lambda_{s,ij},\quad \lambda_{s,ij}=-\frac{m_s}{2}\int\;\dd{\bf c}\;
c^2c_i {\cal C}_{s,j}({\bf c}).
\end{equation}
Upon writing Eqs.\ (\ref{3.24})--(\ref{3.33}) use has been made of the symmetry
properties of ${\boldsymbol {\cal A}}_{r}$, ${\boldsymbol {\cal B}}_{r}$, ${\boldsymbol {\cal C}}_{r}$ and ${\cal D}_{r,k\ell}$. In general, the set of {\em generalized}
transport coefficients defined above are nonlinear functions of the shear rate, the
coefficients of restitution and the parameters of the mixture (masses, sizes and concentration).

\section{Generalized transport coefficients}
\label{sec4}

This Section is devoted to the evaluation of the generalized transport coefficients
associated with the mass, momentum and heat fluxes. We consider each flux separately.

\subsection{Mass flux}

The constitutive form for the mass flux to first order in spatial gradients is given by Eq.\ (\ref{3.24}). To illustrate with some detail the evaluation of the transport
coefficients of the mass flux, let us consider the diffusion coefficients $D_{ij}$,
defined by Eq.\ (\ref{3.27}). These coefficients can be obtained by multiplying both
sides of Eq.\ (\ref{3.18}) by $m_1 c_j$ and integrating over ${\bf c}$. After some
algebra, one arrives at
\begin{eqnarray}
\label{4.1}
& &\left(\frac{2a}{d p}
P_{xy}^{(0)}+\zeta^{(0)}\right)\left(p\partial_p+T\partial_T\right)D_{ij}-\nu_D D_{ij}-a_{ik}D_{kj}\nonumber\\
& &=
\frac{\rho_1}{\rho}\frac{\partial P_{ij}^{(0)}}{\partial x_1}-\frac{\partial P_{1,ij}^{(0)}}{\partial x_1}
-\left(\frac{2a}{d}\frac{\partial P_{xy}^{(0)}}{\partial x_1}+p
\frac{\partial \zeta^{(0)}}{\partial x_1}\right) \nonumber\\
& & \times \frac{\rho^2 D_{p,ij}}{m_1m_2np}
-\left(\frac{2aT}{d p}\frac{\partial P_{xy}^{(0)}}{\partial x_1}+T
\frac{\partial \zeta^{(0)}}{\partial x_1}\right)\frac{\rho^2 D_{T,ij}}{m_1m_2nT}.\nonumber\\
\end{eqnarray}
In Eq.\ \eqref{4.1} use has been made of the results \cite{GA05}
\begin{equation}
\label{4.2}
\int\; \dd{\bf c}\; m_1 c_i \left({\cal L}_1{\cal A}_{1,j}+ {\cal M}_1{\cal A}_{2,j}\right)=
-\frac{m_1m_2n}{\rho}\nu_D D_{ij},
\end{equation}
\begin{equation}
\label{4.3}
\int\; \dd{\bf c}\; m_1 c_i A_{1,j}=\frac{\rho_1}{\rho}\frac{\partial P_{ij}^{(0)}}{\partial x_1}
-\frac{\partial P_{1,ij}^{(0)}}{\partial x_1},
\end{equation}
where
\begin{equation}
\label{4.4}
\nu_D=\frac{\rho \omega_{12}}{d\rho_2}\mu_{21}(1+\alpha_{12}).
\end{equation}
In the hydrodynamic regime, the diffusion tensor can be written as $D_{ij}=D_0D_{ij}^*$ where $D_0=(\rho T/m_1m_2\nu_0)$ and $D_{ij}^*$ is a dimensionless function of the
reduced shear rate $a^*$, the coefficients of restitution $\alpha_{rs}$, the mass ratio $\mu\equiv m_1/m_2$, the ratio of diameters $\sigma_1/\sigma_2$ and the mole
fraction $x_1$. The dependence of $D_{ij}^*$ on the pressure and temperature is through the reduced shear rate $a^*\propto T^{1-\beta}/p$. Thus,
\beqa
\label{4.5}
\left(p\partial_p+T\partial_T\right)D_{ij}&=&\left(p\partial_p+T\partial_T\right)
D_0D_{ij}^*\nonumber\\
&=&(1-\beta)D_{ij}-
\beta D_{0}a^*\frac{\partial D_{ij}^*}{\partial a^*}.\nonumber\\
\eeqa
Consequently, in dimensionless form, Eq.\ (\ref{4.1}) yields
\begin{eqnarray}
\label{4.6} & & \left(\frac{2a^*}{d} P_{xy}^{*}+\zeta^{*}\right)
\left[(1-\beta)D_{ij}^*-\beta a^*\frac{\partial D_{ij}^*}{\partial a^*}\right]-\nu_D^*D_{ij}^*\nonumber\\
& & -a_{ik}^*D_{kj}^*=
\frac{\rho_1}{\rho}\frac{\partial P_{ij}^{*}}{\partial x_1}-\frac{\partial P_{1,ij}^{*}}{\partial x_1}
-\left(\frac{2a^*}{d}\frac{\partial P_{xy}^{*}}{\partial x_1}+
\frac{\partial \zeta^{*}}{\partial x_1}\right)\nonumber\\
& & \times D_{p,ij}^*
-\left(\frac{2a^*}{d}\frac{\partial P_{xy}^{*}}{\partial x_1}+
\frac{\partial \zeta^{*}}{\partial x_1}\right)D_{T,ij}^*.
\end{eqnarray}
Here, $\zeta^*\equiv \zeta^{(0)}/\nu_0$, $P_{r,ij}^*\equiv P_{r,ij}^{(0))}/p$,
$\nu_D^*\equiv \nu_D/\nu_0$, $a_{ij}^*\equiv a_{ij}/\nu_0$, $D_{p,ij}^*\equiv D_{p,ij}/D_{p0}$, and $D_{T,ij}^*\equiv D_{T,ij}/D_{p0}$
where $D_{p0}=(p/\rho \nu_0)$ and we recall that $a_{ij}=a\delta_{ix}\delta_{jy}$. It must be noted that $P_{r,ij}^*$ and $\zeta^*$ depend also on $x_1$ and $a^*$ through their dependence on the temperature ratio $\gamma\equiv T_1/T_2$.

The equations defining the (scaled) tensors $D_{p,ij}^*$ and $D_{T,ij}^*$ can be
obtained by following similar steps as those made before for $D_{ij}^*$. After
some algebra, the results are
%\begin{widetext}
\begin{eqnarray}
\label{4.7}
& & \beta\left(\frac{2a^*}{d} P_{xy}^{*}+\zeta^{*}\right)
\left(D_{p,ij}^*+a^*\frac{\partial D_{p,ij}^*}{\partial a^*}\right)\nonumber\\
& & -\left[\frac{2a^*}{d}
\left(P_{xy}^*-a^*\frac{\partial P_{xy}^*}{\partial a^*}\right)+2\zeta^*-a^*\frac{\partial \zeta^*}{\partial a^*}\right]
D_{p,ij}^*\nonumber\\
& &+\nu_D^*D_{p,ij}^*
+a_{ik}^*D_{p,kj}^*=-\left(\frac{\rho_1}{\rho}P_{ij}^*-P_{1,ij}^*\right)\nonumber\\
& &
+a^*\left(\frac{\rho_1}{\rho}\frac{\partial P_{ij}^*}{\partial a^*}-\frac{\partial P_{1,ij}^*}{\partial a^*}\right)\nonumber\\
& & -
\left(\frac{2}{d}a^{*2}\frac{\partial P_{xy}^*}{\partial a^*}-\zeta^*+a^*\frac{\partial \zeta^*}{\partial a^*}
\right)D_{T,ij}^*,
\end{eqnarray}
\begin{eqnarray}
\label{4.8}
& & \beta\left(\frac{2a^*}{d} P_{xy}^{*}+\zeta^{*}\right)
\left(D_{T,ij}^*+a^*\frac{\partial D_{T,ij}^*}{\partial a^*}\right)-\left[\frac{2a^*}{d}\right.\nonumber\\
& & \left. \times
\left(P_{xy}^*+ (1-\beta)a^*\frac{\partial P_{xy}^*}{\partial a^*}\right)+\beta \zeta^*
+(1-\beta)a^*\frac{\partial \zeta^*}{\partial a^*}\right]\nonumber\\
& & \times D_{T,ij}^*+\nu_D^*D_{T,ij}^*+a_{ik}^*D_{T,kj}^*=-(1-\beta)a^*\left(\frac{\rho_1}{\rho}\frac{\partial P_{ij}^*}{\partial a^*}\right.\nonumber\\
& & \left.
-\frac{\partial P_{1,ij}^*}{\partial a^*}\right)+
(1-\beta)\left(\frac{2}{d}a^{*2}\frac{\partial P_{xy}^*}{\partial a^*}-\zeta^*+a^*\frac{\partial \zeta^*}{\partial a^*}
\right)D_{p,ij}^*.\nonumber\\
\end{eqnarray}
%\end{widetext}
Upon writing Eqs.\ (\ref{4.7}) and (\ref{4.8}) use has been made of the identities
\begin{subequations}
\begin{equation}
\label{4.9} p\partial_pP_{ij}^{(0)}=p\left(P_{ij}^*-a^*\partial_{a^*}P_{ij}^*\right),
\eeq
\beq
T\partial_TP_{ij}^{(0)}=p(1-\beta)a^*\partial_{a^*}P_{ij}^*,
\end{equation}
\end{subequations}
\begin{subequations}
\begin{equation}
\label{4.9.1}
p\partial_p\zeta^{(0)}=\zeta^{(0)}-\nu_0a^*\partial_{a^*}\zeta^*,
\eeq
\beq
T\partial_T \zeta^{(0)}=(\beta-1)\zeta^{(0)}+(1-\beta)\nu_0a^*\partial_{a^*}\zeta^*.
\end{equation}
\end{subequations}
In the absence of shear field ($a^*=0$), $P_{ij}^*=\delta_{ij}$, $P_{r,ij}^*=x_r \gamma_r\delta_{ij}$,
and Eqs.\ (\ref{4.6})--(\ref{4.8}) have the solutions $D_{ij}^*=D^*\delta_{ij}$, $D_{p,ij}^*=D_p^*\delta_{ij}$ and $D_{T,ij}^*=D_T^*\delta_{ij}$ where
\begin{equation}
\label{4.10}
D^*=\left(\nu_D^*-(1-\beta)\zeta^*\right)^{-1}\left[\gamma_1+x_1\frac{\partial \gamma_1}{\partial x_1}+
\frac{\partial \zeta^{*}}{\partial x_1}\left(D_p^*+D_T^*\right)\right],
\end{equation}
\begin{equation}
\label{4.11}
D_p^*=x_1\gamma_1\left(1-\frac{pm_1}{\rho T_1}\right)\left(\nu_D^*+(\beta-2)\zeta^*-(\beta-1)
\frac{\zeta^{*2}}{\nu_D^*}\right)^{-1},
\end{equation}
\begin{equation}
\label{4.12}
D_T^*=(\beta-1)\frac{\zeta^*}{\nu_D^*}D_p^*.
\end{equation}
Equations (\ref{4.10})--(\ref{4.12}) agree with the expressions derived for IMM (model B with $\beta=\frac{1}{2}$) in the Navier-Stokes hydrodynamic order \cite{GA05}.
Beyond the Navier-Stokes domain (vanishing shear rates), in general Eqs.\ (\ref{4.6})--(\ref{4.8}) are nonlinear differential equations that must be solved with
the appropriate boundary conditions. However, in the case of model A ($\beta=0$), Eqs.\ (\ref{4.6})--(\ref{4.8}) become a set of coupled algebraic equations that can be readily solved.

\subsection{Pressure tensor}

The pressure tensor is defined by Eq.\ (\ref{3.25}) in terms of the coefficients
$\eta_{s,ijk\ell}$, Eq.\ (\ref{3.30}). These coefficients can be obtained from the
integral equation (\ref{3.21}) after multiplying it by $m_r c_ic_j$ and integrating
over velocity. The result is
\begin{widetext}
\begin{eqnarray}
\label{4.13} & & \left(\frac{2a}{d p} P_{xy}^{(0)}+\zeta^{(0)}\right)
(p\partial_p+T\partial_T)\eta_{1,ijk\ell}-\left(a_{ip}\eta_{1,jpk\ell}+a_{jp}
\eta_{1,ipk\ell}-a_{pk}\eta_{1,ijp\ell}\right)
-\left(\tau_{11}\eta_{1,ijk\ell}+\tau_{12}\eta_{2,ijk\ell}\right)\nonumber\\
& &=p\frac{\partial P_{1,ij}^{(0)}}{\partial p}\delta_{k\ell}-
\left(\delta_{k\ell}P_{1,ij}^{(0)}+\delta_{ik}P_{1,j\ell}^{(0)}+\delta_{jk}P_{1,i\ell}^{(0)}\right)
 +\frac{2}{d
p}\left(P_{k\ell}^{(0)}-a\eta_{xyk\ell}\right)\left(p\frac{\partial P_{1,ij}^{(0)}}{\partial p}+
T\frac{\partial P_{1,ij}^{(0)}}{\partial T}\right).
\end{eqnarray}
\end{widetext}
The corresponding equation for $\eta_{2,ijk\ell}$ can be obtained from Eq.\ (\ref{4.13}) by changing $1\leftrightarrow 2$.
Upon writing (\ref{4.13}) use has been made of the result \cite{GA05}
\beq
\label{4.14}
\int \dd{\bf c}m_1 c_ic_j \left({\cal L}_1{\cal D}_{1,k\ell}+ {\cal M}_1{\cal D}_{2,k\ell}\right)
=-\tau_{11}\eta_{1,ijk\ell}-\tau_{12}\eta_{2,ijk\ell},
%\nonumber\\
\eeq
where
\beqa
\label{4.15}
\tau_{11}&=&\frac{\omega_{11}}{d(d+2)}(1+\alpha_{11})(d+1-\alpha_{11})\nonumber\\
& + & 2\frac{\omega_{12}}{d}\mu_{21}
(1+\alpha_{12})\left[1-\frac{\mu_{21}(1+\alpha_{12})}{d+2}\right],
\eeqa
\begin{equation}
\label{4.16}
\tau_{12}=-2\frac{\omega_{12}}{d(d+2)}\frac{\rho_1}{\rho_2}\mu_{21}^2
(1+\alpha_{12})^2.
\end{equation}
The coefficients $\eta_{r,ijk\ell}$ can be written as
$\eta_{r,ijk\ell}=(p/\nu_0)\eta_{r,ijk\ell}^*$. The dependence of
$\eta_{r,ijk\ell}^*$ on $p$ and $T$ is through $a^*$ so that
\beqa
\label{4.17}
(p\partial_p+T\partial_T)\eta_{r,ijk\ell}&=&\left(p\partial_p+T\partial_T\right) \frac{p}{\nu_0}\eta_{r,ijk\ell}^*\nonumber\\
&=&(1-\beta)\eta_{r,ijk\ell}-\frac{\beta p}{\nu_0} a^*\frac{\partial\eta_{r,ijk\ell}^*}{\partial a^*}. \nonumber\\
\eeqa
Thus, in dimensionless form, Eq.\ (\ref{4.13}) can be finally written as
\begin{widetext}
\begin{eqnarray}
\label{4.18} & & \left(\frac{2}{d}a^* P_{xy}^{*}+\zeta^*\right)
\left[(1-\beta)\eta_{1,ijk\ell}^*-\beta a^*\frac{\partial \eta_{1,ijk\ell}^*}{\partial a^*}\right]
-\left(a_{ip}^*\eta_{1,jpk\ell}^*+a_{jp}^*
\eta_{1,ipk\ell}^*-a_{pk}^*\eta_{1,ijp\ell}^*\right)
-\left(\tau_{11}^*\eta_{1,ijk\ell}^*+\tau_{12}^*\eta_{2,ijk\ell}^*\right)\nonumber\\
& &=-a^*\frac{\partial P_{1,ij}^*}{\partial a^*}\delta_{k\ell}-
\left(\delta_{ik}P_{1,j\ell}^{*}+\delta_{jk}P_{1,i\ell}^{*}\right)
+\frac{2}{d}\left(P_{k\ell}^{*}-a^*\eta_{xyk\ell}^*\right)\left(P_{1,ij}^{*}-\beta a^*
\frac{\partial P_{1,ij}^*}{\partial a^*}\right),
\end{eqnarray}
\end{widetext}
where $\tau_{ij}^*\equiv \tau_{ij}/\nu_0$.

In the case of mechanically equivalent particles, $P_{1,ij}^*/x_1=P_{2,ij}^*/x_2=P_{ij}^*$,
$\eta_{1,ijk\ell}^*/x_1=\eta_{2,ijk\ell}^*/x_2=\eta_{ijk\ell}^*$, where $\eta_{ijk\ell}^*$ verifies the differential equation
\beqa
\label{4.18.1}
& &\left(\frac{2}{d}a^* P_{xy}^{*}+\zeta^*\right)
\left[(1-\beta)\eta_{ijk\ell}^*-\beta a^*\frac{\partial \eta_{ijk\ell}^*}{\partial a^*}\right]\nonumber\\
& &
-\left(a_{ip}^*\eta_{jpk\ell}^*+a_{jp}^*
\eta_{ipk\ell}^*-a_{pk}^*\eta_{ijp\ell}^*\right)-\nu_\eta^* \eta_{ijk\ell}^*
\nonumber\\
& &=-a^*\frac{\partial P_{ij}^*}{\partial a^*}\delta_{k\ell}-
\left(\delta_{ik}P_{j\ell}^{*}+\delta_{jk}P_{i\ell}^{*}\right)
\nonumber\\
& &
+\frac{2}{d}\left(P_{k\ell}^{*}-a^*\eta_{xyk\ell}^*\right)\left(P_{ij}^{*}-\beta a^*
\frac{\partial P_{ij}^*}{\partial a^*}\right),
\eeqa
where
\beq
\label{4.18.2}
\nu_\eta^*=\frac{(1+\al)(d+1-\al)}{d(d+2)}.
\eeq
Equation \eqref{4.18.1} agrees with the results derived in Ref.\ \cite{G07} for a sheared monocomponent granular gas of IMM. In the limit of vanishing shear rates ($a^*=0$),
the solution to Eq.\ (\ref{4.18}) can be written as
\begin{equation}
\label{4.19}
\eta_{ijk\ell}^*=(\eta_{1}^*+\eta_{2}^*) \Delta_{ijk\ell}, \quad \Delta_{ijk\ell}
=\delta_{ik}\delta_{j\ell}+\delta_{jk}\delta_{i\ell}-
\frac{2}{d}\delta_{ij}\delta_{k\ell},
\end{equation}
where
\begin{equation}
\label{4.20a}
\eta_{1}^*=\frac{x_1\gamma_1[\tau_{22}^*-(1-\beta)\zeta^*]-x_2\gamma_2\tau_{12}}{[\tau_{11}-(1-\beta)\zeta^*]
[\tau_{22}-(1-\beta)\zeta^*]-\tau_{12}\tau_{21}},
\end{equation}
\begin{equation}
\label{4.20b}
\eta_{2}^*=\frac{x_2\gamma_2[\tau_{11}^*-(1-\beta)\zeta^*]-x_1\gamma_1\tau_{21}}
{[\tau_{11}-(1-\beta)\zeta^*][\tau_{22}-(1-\beta)\zeta^*]-\tau_{12}\tau_{21}}.
\end{equation}
For model B with $\beta=1/2$, Eqs.\ \eqref{4.19}--\eqref{4.20b} are consistent with those previously obtained for the Navier-Stokes shear viscosity of an inelastic binary Maxwell
mixture \cite{GA05}. On the other hand, except in the above two limit cases,  Eq.\ (\ref{4.18}) for $\eta_{1,ijk\ell}^*$  and its counterpart for $\eta_{2,ijk\ell}^*$
can be only solved analytically for model A ($\beta=0$).

The evaluation of the transport coefficients associated with the heat flux is more involved than the one carried out before for the mass and momentum fluxes. For the sake of
brevity, only the final expressions of the differential equations defining the coefficients $D_{ij}''$, $L_{ij}$ and $\lambda_{ij}$ are provided (see Appendix \ref{appB}).

\section{Tracer limit}
\label{sec5}

The results obtained in the preceding Section apply for models A and B and give all the relevant information on the
influence of shear flow on the mass, momentum and heat transport of a granular binary mixture. According to these results, the set of generalized (dimensionless)
transport coefficients $\{D_{ij}^*, D_{p,ij}^*, D_{T,ij}^*,
\eta_{ijk\ell}^*,\ldots \}$ are nonlinear functions of the
(reduced) shear rate, the concentration $x_1$ and the mechanical parameters of the mixture (mass and size ratios and
coefficients of restitution) without any restriction on their values.
\vicente{On the other hand, the evaluation of these coefficients (even in the case of model A where the results are analytic) is quite tedious due essentially to the complex dependence of the partial pressure tensors ${\sf P}_{r,ij}^{(0)}$ and the temperature ratio $\gamma$ on both the mole
fraction $x_1$ and the (reduced) shear rate $a^*$. Thus, for the sake of simplicity, we consider the tracer limit  ($x_1\to 0$) where   the mass flux is the relevant flux since the momentum and heat fluxes of the system (intruder plus gas
particles) are the same as those previously obtained \cite{G07} for a monocomponent granular gas of IMM.}

It must be remarked that a non-equilibrium phase transition has been recently \cite{GT11} identified in the tracer limit for a granular binary mixture of IMM.
This transition refers to the existence
of a region (coined as the ordered phase) where the contribution of tracer particles to the total kinetic energy of the system is \emph{finite}.
However, the above (surprising) behavior has been
only analytically found when the collision frequency $\omega_{rs}$ is assumed to be independent of the temperature ratio (``plain vanilla Maxwell model'')
and hence, it does not seem to exist for the
more realistic version of the IMM considered here. The effects of the above transition on the Navier-Stokes transport coefficients of a granular binary mixture has been recently studied \cite{GKT15}.

\subsection{Model A}
\label{sub5a}

In the tracer limit, ${\sf P}^{(0)}\simeq {\sf P}_2^{(0)}$ and the relevant elements of the partial pressure tensor ${\sf P}_1^{(0)}$  admit simplified forms (see Appendix \ref{appC}).
In particular, in the tracer limit, $\gamma_1\simeq \gamma$, $\partial_{x_1}P_{ij}^{(0)}=\partial_{x_1}\gamma=\partial_{x_1}\zeta^*=\partial_{a^*}\zeta^*=0$ and $\partial_{x_1}P_{1,ij}^{(0)}=P_{1,ij}^{(0)}/x_1$. Taking
into account these
simplifications, and for model A ($\beta=0$), Eqs.\ (\ref{4.7})--(\ref{4.8}) become
\begin{equation}
\label{5.1} \left(\frac{2a^*}{d}
P_{2,xy}^{*}+\zeta^{*}\right)D_{ij}^*-\nu_D^*D_{ij}^*-a_{ik}^*D_{kj}^*=
-x_1^{-1}P_{1,ij}^*,
\end{equation}
\begin{eqnarray}
\label{5.2}
& & \left[\frac{2a^*}{d}
\left(P_{2,xy}^*-a^*\frac{\partial P_{2,xy}^*}{\partial a^*}\right)+2\zeta^*\right] D_{p,ij}^*
-\nu_D^*D_{p,ij}^*
\nonumber\\
& &
-a_{ik}^*D_{p,kj}^*=x_1 \mu P_{2,ij}^*- P_{1,ij}^* -x_1 \mu \; a^*\frac{\partial P_{2,ij}^*}{\partial a^*}\nonumber\\
& & +a^*\frac{\partial P_{1,ij}^*}{\partial a^*}
+\left(\frac{2a^{*2}}{d}\frac{\partial P_{2,xy}^*}{\partial a^*}-
\zeta^{*}\right)D_{T,ij}^*,
\end{eqnarray}
\begin{eqnarray}
\label{5.3}
& &\frac{2a^*}{d}\left(P_{2,xy}^* +a^*\frac{\partial P_{2,xy}^*}{\partial a^*}\right)D_{T,ij}^*
-\nu_D^*D_{T,ij}^*-a_{ik}^*D_{T,kj}^*
\nonumber\\
& &
=x_1 \mu \; a^*\frac{\partial P_{2,ij}^*}{\partial a^*}-a^*\frac{\partial P_{1,ij}^*}{\partial a^*}
 -\left(\frac{2a^{*2}}{d}\frac{\partial P_{2,xy}^*}{\partial a^*}-
\zeta^{*}\right)D_{p,ij}^*.
\nonumber\\
\end{eqnarray}
Here, $\zeta^*=(1-\al_{22}^2)/2d$ and $\nu_D^*=(\omega_{12}^*\mu_{21}(1+\al_{12}))/d$ where $\omega_{12}^*\equiv \omega_{12}/\nu_0$. Upon deriving Eq.\ \eqref{5.1} we have neglected the
contributions coming from the tensors $D_{p,ij}^*$ and $D_{T,ij}^*$ since both tensors are proportional to $x_1$ (and hence, they vanish in the tracer limit) while $D_{ij}^*$ is
independent of $x_1$. In addition, the derivatives $\partial_{a^*}P_{2,ij}^{*}$ and $\partial_{a^*}P_{1,ij}^{*}$ appearing in Eqs.\ \eqref{5.1}--\eqref{5.3} are obtained in
Appendix \ref{appD}.
\begin{figure}
%[hbtp]
%\begin{center}
%\begin{tabular}{lr}
%\resizebox{3.7cm}{!}
{\includegraphics[width=0.75\columnwidth]{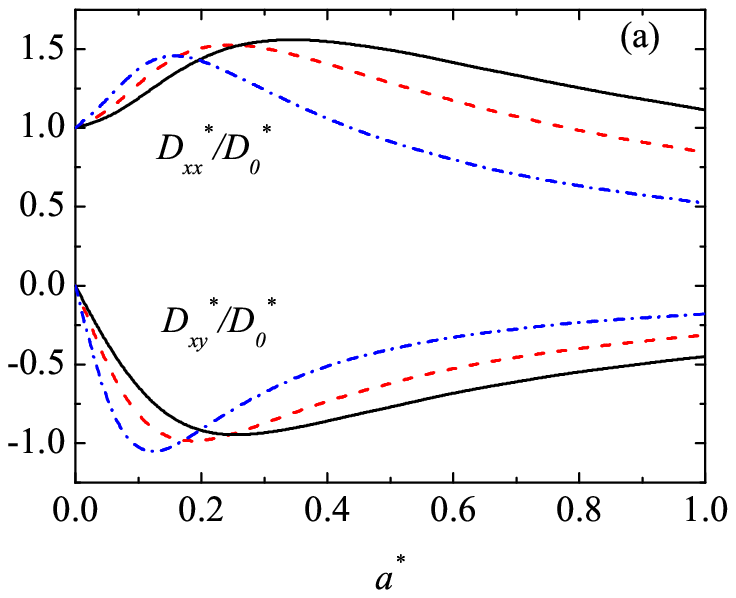}}
%&\resizebox{3.2cm}{!}
{\includegraphics[width=0.75\columnwidth]{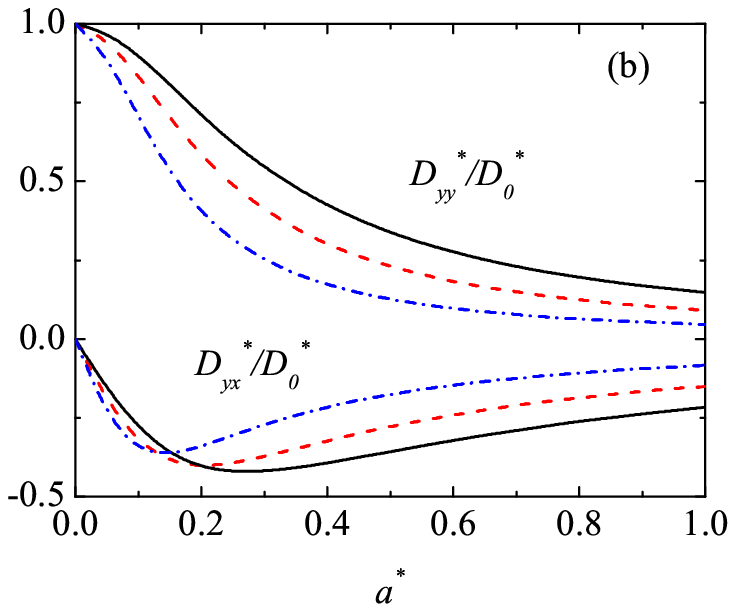}}
%\end{tabular}
%\resizebox{6.5cm}{!}{\includegraphics{fig4.eps}}
%\end{center}
\caption{(Color online) Panel (a): Shear-rate dependence of the (dimensionless) coefficients $D_{xx}^*/D_0^*$ and $D_{xy}^*/D_0^*$ for $d=3$, $\sigma_1/\sigma_2=1$, $m_1/m_2=2$
and three different values of the (common) coefficient of restitution $\al$: $\al=1$ (solid line), $\al=0.8$ (dashed red line) and $\al=0.6$ (dash-dotted blue line). Panel (b):
Shear-rate dependence of the (dimensionless) coefficients $D_{yy}^*/D_0^*$ and $D_{yx}^*/D_0^*$ for the same parameter values. These results pertain to model A ($\beta=0$).
\label{fig1}}
\end{figure}

As in the case of IHS \cite{G07bis}, the coefficients $D_{ij}$ decouple from the other ones and hence,
they can be obtained straightforwardly. Their expressions are
\beq
\label{5.4}
D_{ij}^*=\frac{x_1^{-1}}{\nu_D^*-\frac{2a^*}{d}P_{2,xy}^*-\zeta^*}\left(P_{1,ij}^*-\frac{a_{ik}^*P_{1,kj}^*}
{\nu_D^*-\frac{2a^*}{d}P_{2,xy}^*-\zeta^*}\right).
\eeq
\vicente{In the steady state ($(2a^*/d)P_{2,xy}^*+\zeta^*=0$), Eq.\ \eqref{5.4} is consistent with previous results derived for IMM under shear flow \cite{G03}.} The remaining coefficients $D_{p,ij}^*$ and $D_{T,ij}^*$ are coupled and they obey the set of simple algebraic equations \eqref{5.2}--\eqref{5.3}. As alluded to
above, they are proportional to the concentration $x_1$ and thus vanish in the tracer limit. Yet, it is of interest to normalize them by
their vanishing shear rate counterparts (which are also proportional to $x_1$), to study their dependence on parameters other than $x_1$.
In order to illustrate the
shear-rate dependence of the set of transport coefficients $\Delta_{ij}\equiv \left\{D_{ij}^*, D_{p,ij}^*, D_{T,ij}^* \right\}$, we consider a three-dimensional ($d=3$) granular mixture.
Also, to reduce the number of independent parameters, the simplest case of a common coefficient of restitution ($\alpha\equiv \alpha_{22}=\al_{12}$) is studied. Thus, the parameter space
is reduced to four quantities $\left\{\sigma_1/\sigma_2, m_1/m_2, \al, a^*\right\}$.
\begin{figure}
%[hbtp]
%\begin{center}
%\begin{tabular}{lr}
%\resizebox{3.7cm}{!}
{\includegraphics[width=0.75\columnwidth]{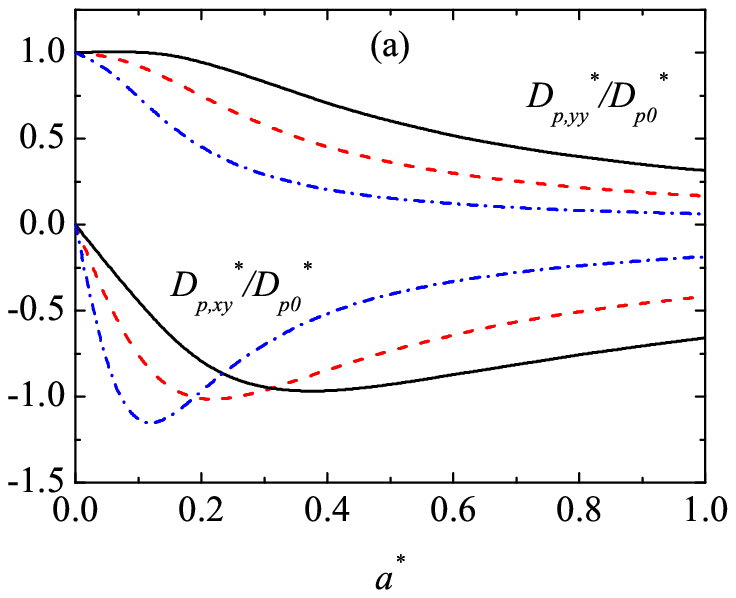}}
%&\resizebox{3.2cm}{!}
{\includegraphics[width=0.75\columnwidth]{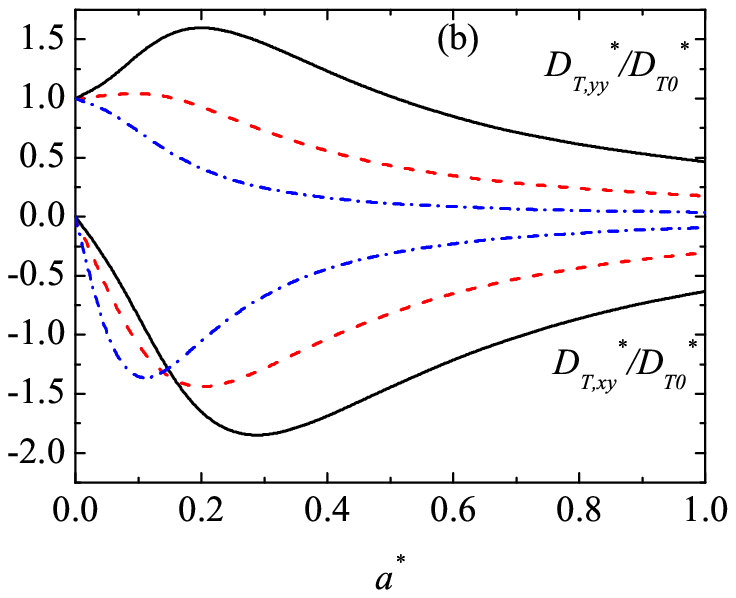}}
%\end{tabular}
%\resizebox{6.5cm}{!}{\includegraphics{fig4.eps}}
%\end{center}
\caption{(Color online) Same as Fig. \ref{fig1} for $D_{p,yy}^*/D_{p,0}^*$ and $D_{p,xy}^*/D_{p,0}^*$ (panel (a)), $D_{T,yy}^*/D_{T,0}^*$ and $D_{T,xy}^*/D_{T,0}^*$
(panel (b)).
\label{fig2}}
\end{figure}
\begin{figure}
\includegraphics[width=0.75 \columnwidth,angle=0]{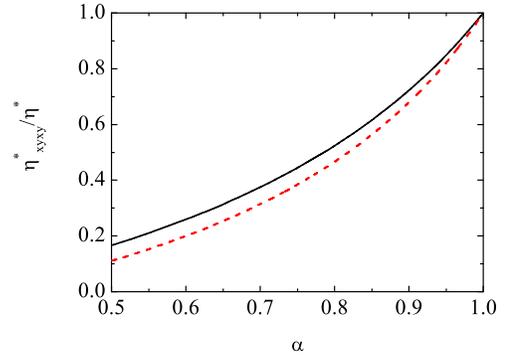}
\caption{(Color online) Plot of the ratio $\eta_{xyxy}^*/\eta^*$ as a function of the coefficient of restitution $\al$ in the
steady state for a monodisperse granular gas (IMM with $\beta=0$, model A). The solid line is the result for a three-dimensional system ($d=3$)
while the dashed line corresponds to a two-dimensional system ($d=2$).
\label{fig7}}
\end{figure}

According to Eqs.\ \eqref{5.1}--\eqref{5.3}, we have that $\Delta_{xz}=\Delta_{zx}=\Delta_{yz}=\Delta_{zy}=0$ in agreement with the symmetry of the linear shear flow \eqref{2.11}.
Thus, there are five nonzero elements of the tensors $\Delta_{ij}$: the three diagonal ($\Delta_{xx}$, $\Delta_{yy}$, and $\Delta_{zz}$) and the two off-diagonal elements ($\Delta_{xy}$ and
$\Delta_{yx}$). The algebraic equations \eqref{5.1}--\eqref{5.3} also show that the anisotropy induced by the shear flow yields the properties $\Delta_{xx}\neq \Delta_{yy} = \Delta_{zz}$
and  $\Delta_{xy}\neq \Delta_{yx}$. The equality $P_{1,yy}^*=P_{1,zz}^*$ implies $\Delta_{yy}=\Delta_{zz}$. This is a consequence of the interaction model considered since
$\Delta_{yy} \neq \Delta_{zz}$  for IHS \cite{G07bis}.

The shear-rate dependence of the relevant elements of the diffusion tensors $D_{ij}^*$, $D_{p,ij}^*$ and $D_{T,ij}^*$ has been plotted in Figs.\ \ref{fig1} and \ref{fig2} for $d=3$,
$\sigma_1/\sigma_2=1$, $m_1/m_2=2$ and three different values of the (common) coefficient of restitution. Here, the tensors have been reduced with respect to their values at zero shear rate,
namely, $D_{ij}^*/D_0^*$, $D_{p,ij}^*/D_{p,0}^*$ and $D_{T,ij}^*/D_{T,0}^*$ where
\beq
\label{5.6}
D_0^*=\frac{\gamma}{\nu_D^*-\zeta^*},\quad D_{T0}^*=-\frac{\zeta^*}{\nu_D^*}D_{p0}^*,
\eeq
\beq
\label{5.6.1}
D_{p0}^*=x_1 \gamma \left(1-\frac{\mu}{\gamma}\right)\left(\nu_D^*-2\zeta^*
+\frac{\zeta^{*2}}{\nu_D^*}\right)^{-1}.
\eeq
It can be seen that the influence of the shear flow on the diffusion coefficients is in general quite important. We also observe that the anisotropy of the system, as measured by
the difference $D_{xx}^*-D_{yy}^*$ grows with both the shear rate and collisional dissipation. As expected, the shear field induces cross effects in the diffusion of particles.
This is measured by the off-diagonal elements $D_{xy}^*$ ($D_{yx}^*$), $D_{p,xy}^*$ ($D_{p,yx}^*$) and $D_{T,xy}^*$ ($D_{T,yx}^*$). These coefficients give the mass transport
along the $x$ ($y$) axis due to spatial gradients parallel to $y$ ($x$) axis. All these coefficients are negative in the region of parameter space explored. We see that,
regardless of the value of $\al$, the shapes of the off-diagonal elements are quite similar: there is a region of values of $a^*$ for which their magnitude increase with increasing shear rate,
while the opposite happens for larger shear rates. With respect to the diagonal elements, they are monotonically decreasing functions of the shear rate (shear-thinning effect),
except in the region of small shear rates. In addition, Figs.\ \ref{fig1} and \ref{fig2} also show that, at a given value of $a^*$, their values decrease with dissipation.

It is also interesting to weigh the respective importance of the zeroth- and first-order
contributions to the (nonlinear) shear viscosity. The zeroth-order USF viscosity, $\eta^*$, is defined by
Eq.\ \eqref{2.12.2} while its (dimensionless) first-order contribution is given by the coefficient $\eta_{xyxy}^*$.
In the steady state and for mechanically equivalent particles, the ratio $\eta_{xyxy}^*/\eta^*$ can be obtained from
Eq.\ \eqref{4.18.1} for model A ($\beta=0$):
\beq
\label{5.6.2}
\frac{\eta_{xyxy}^*}{\eta^*}=\frac{(1+\al)^2}{2(d+2)}\frac{1+2\Lambda_s}{(\nu_{\eta}^*-\zeta^*)(1+6\Lambda_s)},
\eeq
where $\nu_{\eta}^*$ is defined by Eq.\ \eqref{4.18.2} and $\Lambda_s=\frac{(d+2)(1-\al)}{2d(1+\al)}$. Figure \ref{fig7} shows the ratio $\eta_{xyxy}^*/\eta^*$ versus the
coefficient of restitution $\al$ for spheres ($d=3$) and disks ($d=2$). For elastic collisions ($\al=1$), $a^*=0$ in the steady state and so, $\eta_{xyxy}^*=\eta^*$.
Moreover, the zeroth-order solution $\eta^*$ generically gives a significant
%non negligible
contribution to the total non-Newtonian shear viscosity.
%since the ratio $\eta_{xyxy}^*/\eta^*$ decreases with dissipation.
Figure \ref{fig7} also displays a pronounced shear thinning effect:
when $\alpha$ decreases, the
steady state gas departs more and more from equilibrium, and the resulting $a^*$ increases;
this in turn leads to a decrease of $\eta_{xyxy}^*/\eta^*$. Thus, the shear thinning effect
is more marked for $\eta_{xyxy}^*$ than for $\eta^*$. Indeed, Ref. \cite{GT10}
has shown that the USF viscosity itself, $\eta^*$, exhibits shear thinning.
For further details dealing with the detailed behavior of $\eta^*$, see Ref. \cite{GT10}.

\begin{figure}
%[hbtp]
%\begin{center}
%\begin{tabular}{lr}
%\resizebox{3.7cm}{!}
{\includegraphics[width=0.75\columnwidth]{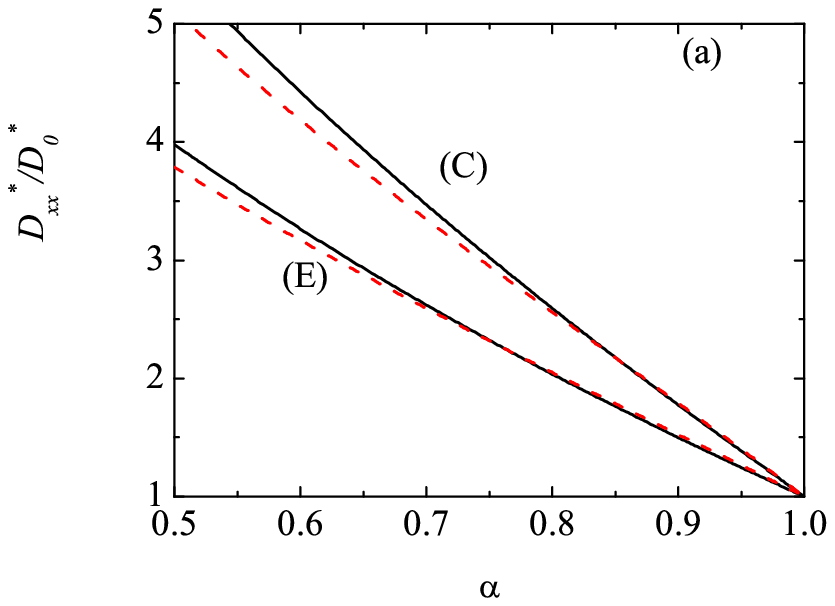}}
%&\resizebox{3.2cm}{!}
{\includegraphics[width=0.77\columnwidth]{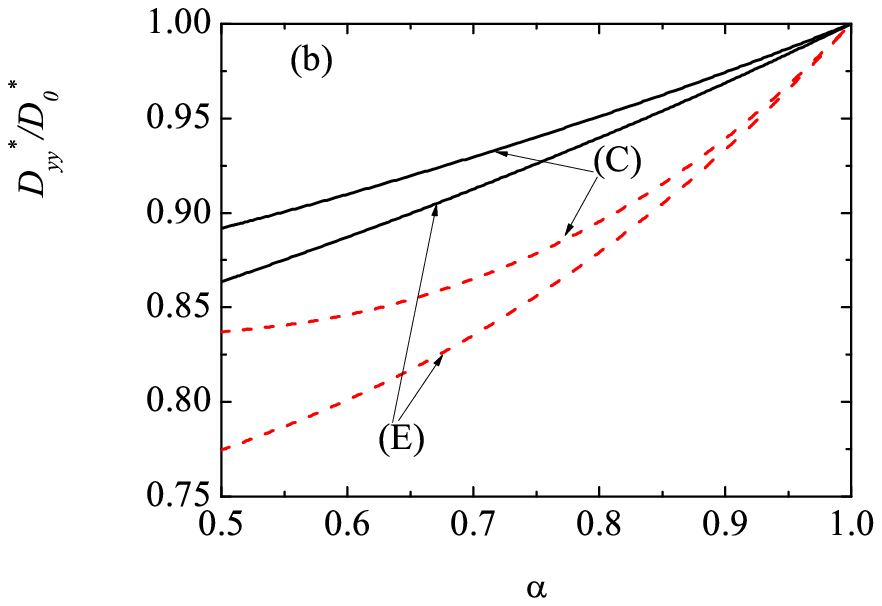}}
%\end{tabular}
%\resizebox{6.5cm}{!}{\includegraphics{fig4.eps}}
%\end{center}
\caption{(Color online) Plot of the diagonal (dimensionless) coefficients $D_{xx}^*/D_0^*$ (panel (a)) and $D_{yy}^*/D_0^*$ (panel (b)) as functions of the (common) coefficient of
restitution $\al$ in the steady USF state for $d=3$ in the cases $\sigma_1/\sigma_2=1$ and $m_1/m_2=2$ (C) and $\sigma_1/\sigma_2=2$ and $m_1/m_2=4$ (E). The solid lines correspond to the results derived
here for IMM (models A and B) while the dashed lines are the results obtained for IHS \cite{G02,G07bis}.
\label{fig3}}
\end{figure}
\begin{figure}
%[hbtp]
%\begin{center}
%\begin{tabular}{lr}
%\resizebox{3.7cm}{!}
{\includegraphics[width=0.75\columnwidth]{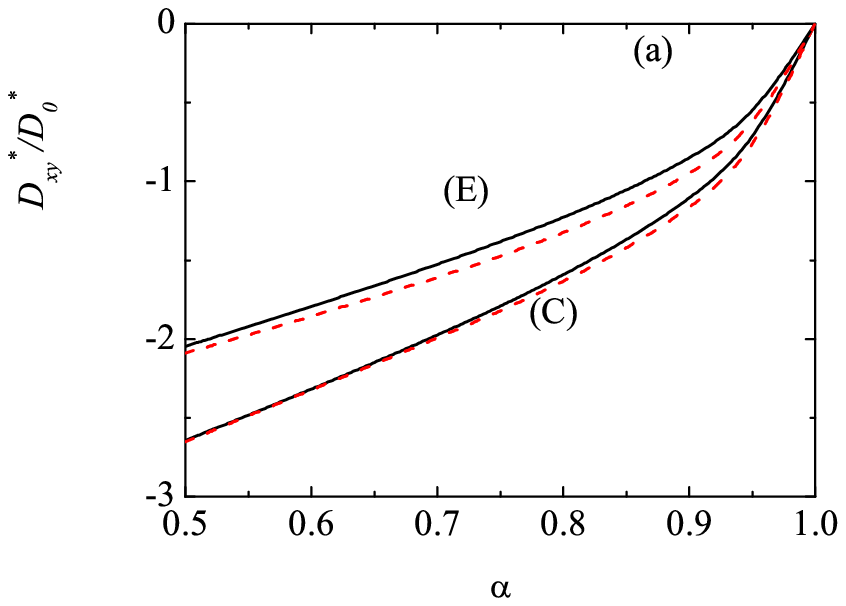}}
%&\resizebox{3.2cm}{!}
{\includegraphics[width=0.77\columnwidth]{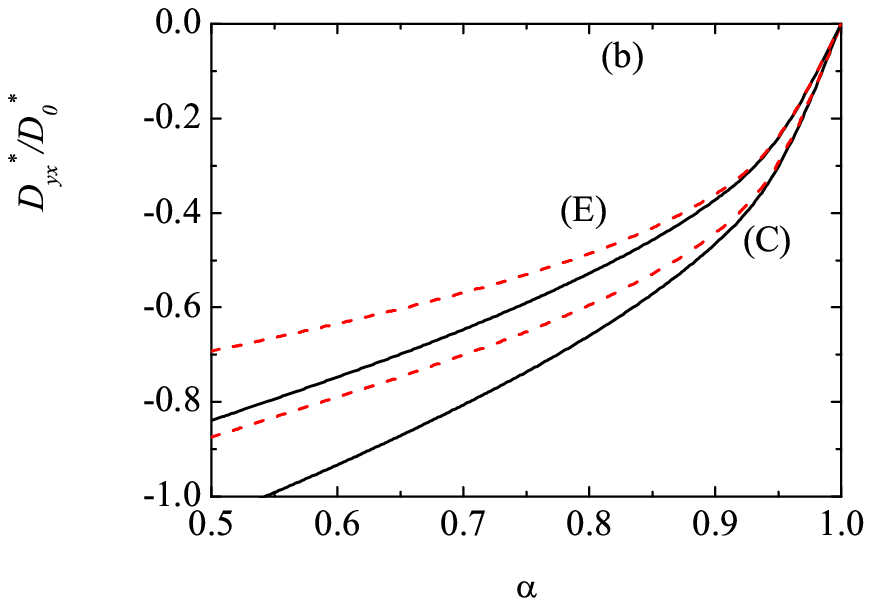}}
%\end{tabular}
%\resizebox{6.5cm}{!}{\includegraphics{fig4.eps}}
%\end{center}
\caption{(Color online) Same as Fig. \ref{fig3} for
$D_{xy}^*/D_0^*$ (panel (a)) and $D_{yx}^*/D_0^*$ (panel (b)).
\label{fig4}}
\end{figure}

\subsection{Model B: steady state conditions}
\label{sec5b}

In model B the collision frequency $\nu_0(T)$ is an increasing function of temperature and hence, the (reduced) shear rate $a^*(T)=a/\nu_0(T)$ depends on time. Thus, in order to determine
the diffusion coefficients one would have to solve numerically
Eqs.\ \eqref{4.6}--\eqref{4.8} in the tracer limit, discard the kinetic stage of the evolution and eliminate time in favor of $a^*(t)$ \cite{SGD04,SG07}. An additional technical difficulty
in the case of granular mixtures is that the diffusion coefficients depend also on the temperature ratio, that is itself time dependent through its dependence on $a^*(t)$.
The integration of Eqs.\ \eqref{4.6}--\eqref{4.8} is therefore a significantly more complex problem than for the monodisperse system.
On the other hand, given that the results derived for the rheological properties in a single granular gas under USF \cite{GS07} indicate that the
influence of the temperature dependence on $\nu_0$ on rheology is quite small, one can consider the steady-state solution for model B, which is at any rate of interest in its own right.
In this case, the condition \eqref{2.12.1} applies and the solution to Eqs.\ \eqref{4.6}--\eqref{4.8} can be obtained analytically in the tracer limit.
Here, we focus our attention on the tracer diffusion tensor $D_{ij}^*$ whose expression in the steady state is \emph{universal} since it applies for both models A and B, regardless the specific dependence of $\nu_0$ on $T$.

\vicente{A previous comparison between IMM and IHS for this tensor was carried out in Ref.\ \cite{G03}. However, the (approximate) theoretical results for IHS considered in Fig.\ 7 of \cite{G03} were obtained from a Grad's solution \cite{G49} where the weight distribution is Gaussian \cite{G02} instead of the shear flow distribution (zeroth-order solution) \cite{G07bis}. In the comparison performed here, we will use the latter predictions of IHS \cite{G07bis} which are expected to be more reliable than the other ones \cite{G03}. In this sense, the present comparison complements to the one made before in Ref.\ \cite{G03}.}

As in previous works on IMM \cite{G03,GA05,MGV14} and in order to compare the results between IMM and IHS, the parameter $A$ appearing in the definition of $\nu_0$ (see Eq.\ \eqref{2.5}) is chosen as
\beq
\label{5.9}
A=\frac{\Omega_d}{\sqrt{\pi}}\sigma_{12}^{d-1}\sqrt{\frac{2(m_1+m_2)}{m_1m_2}}.
\eeq
With this choice, the partial cooling rates $\zeta_r$ (associated with the partial temperatures $T_r$) of IMM (with $\beta=1/2$) are the same as those obtained for IHS
(as evaluated in the Maxwellian approximation) \cite{GD99}. The dependence of the set of tracer diffusion coefficients $\left\{D_{xx}^*, D_{yy}^*, D_{xy}^*, D_{yx}^*\right\}$
on the (common) coefficient of restitution $\al$ is illustrated in Figs.\ \ref{fig3} and \ref{fig4} for two different systems. We observe in general a good agreement between
IMM and IHS, especially in the case of the coefficients $D_{xx}^*$ and $D_{xy}^*$. These coefficients measure mass transport in the flow direction ($x$ axis). It must be pointed out that the discrepancies between both interaction models turn out to be more significant as the disparity of masses or sizes increases.

\section{An application: segregation of an intruder by thermal diffusion}
\label{sec6}

As an application of the previous results, this section is devoted to the study of thermal diffusion segregation of an intruder in a sheared granular dilute gas.
Segregation and mixing of dissimilar grains is one of the
most interesting problems in granular mixtures, not only from a fundamental point of view but also from a more practical
perspective. This problem has spawned a number of important experimental, computational, and theoretical works
in the field of granular media, especially when the system is fluidized by vibrating walls \cite{DS13}. In the case of sheared systems, some computational and experimental
works in annular Couette cells \cite{couette} have shown that granular materials segregate by particle size when subjected to shear. On the other hand, in spite of the
relevance of the problem, much less is known on the theoretical description of segregation in sheared granular systems. Previous theoretical studies \cite{AJ04}
on the subject for dense systems have been based on a Chapman-Enskog expansion around Maxwellian distributions with the same temperature for each species. As mentioned before,
the assumption of energy equipartition can be only justified for nearly elastic gases which means small shear rates in the \emph{steady} USF state.

Thermal diffusion is caused by the relative motion of the components of a mixture due to the presence of a temperature
gradient. As a result of this motion, a steady state is finally reached in which the separating effect arising from
thermal diffusion is balanced by the remixing effect of ordinary diffusion \cite{KCL87}. The new feature of our study is to
assess the impact of shear flow on segregation. Under these conditions, the so-called thermal diffusion
factor $\Lambda$ characterizes the amount of segregation parallel to the temperature gradient. However, due to the anisotropy induced by the shear field, a tensor $\boldsymbol{\Lambda}$ rather than a scalar $\Lambda$ is needed to characterize segregation in the different directions. Here, for the sake of simplicity, we consider a situation where the temperature gradient is orthogonal to the shear flow plane (i.e., $\partial_x T=\partial_y T=0, \partial_zT \neq 0$ and $\partial_x u_y=a\equiv \text{const.}$). In this case, the amount of segregation parallel to the thermal gradient is measured by the diffusion factor $\Lambda_z$ defined by the relation
\beq
\label{6.1}
\Lambda_z \frac{\partial \ln T}{\partial z}=-\frac{\partial \ln x_1}{\partial z}.
\eeq
If we assume that the bottom plate is hotter than the top plate ($\partial_z T<0$), then the intruder rises with respect to the gas particles if $\Lambda_z>0$ (i.e., $\partial_z \ln x_1>0$)
while the intruder falls with respect to the gas
particles if $\Lambda_z<0$ (i.e., $\partial_z \ln x_1<0$).

\begin{figure}
\includegraphics[width=0.75 \columnwidth,angle=0]{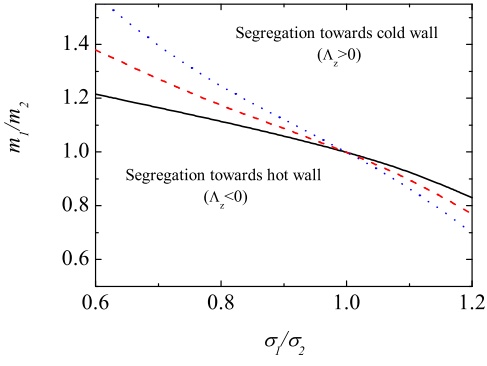}
\caption{(Color online) Phase diagram for segregation for a three-dimensional system ($d=3$) and three different values of the (common) coefficient of restitution
$\alpha\equiv \alpha_{22}=\alpha_{12}$: $\alpha=0.9$ (solid line), $\alpha=0.8$ (dashed line) and $\alpha=0.7$ (dotted line).
\label{fig5}}
\end{figure}
\begin{figure}
\includegraphics[width=0.75 \columnwidth,angle=0]{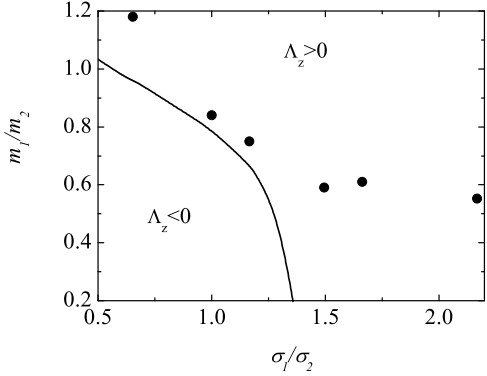}
\caption{(Color online) Phase diagram for segregation for a three-dimensional system ($d=3$) with $\alpha_{22}=0.9$ and $\alpha_{12}=0.7$. The solid line corresponds to the theoretical prediction obtained from Eq.\ \eqref{6.6} while the symbols refer to computer simulations carried out in Ref.\ \cite{VGK14} for IHS in the so-called LTu flow (Couette flow with uniform heat flux).
\label{fig6}}
\end{figure}
Our goal here is to determine $\Lambda_z$ in a steady state with $\delta \mathbf{u}=\mathbf{0}$  and $x_1 \to 0$ (tracer limit) where the spatial gradients of $T$, $p$ and $x_1$ point in the
$z$-direction. Under these conditions, the balance equation \eqref{3.3} yields $j_{1,z}^{(1)}=0$ where $j_{1,z}^{(1)}$ is given by
\beq
\label{6.2}
j_{1,z}^{(1)}=-m_1 D_{zz} \frac{\partial x_1}{\partial z}-\frac{m_2}{T} D_{p,zz} \frac{\partial p}{\partial z}
-\frac{m_2n_2}{T}D_{T,z}\frac{\partial T}{\partial z}.
\eeq
According to Eq.\ \eqref{6.2}, the condition $j_{1,z}^{(1)}=0$ leads to
\beq
\label{6.3}
\frac{\partial x_1}{\partial z}=-\frac{m_2}{m_1 T}\frac{D_{p,zz}}{D_{zz}} \frac{\partial p}{\partial z}-
\frac{m_2n_2}{m_1 T}\frac{D_{T,zz}}{D_{zz}}\frac{\partial T}{\partial z}.
\eeq
In the steady state, the momentum balance equation \eqref{3.4} reduces simply to $\partial_z P_{zz}=0$. The pressure tensor has the form $P_{zz}=p P_{zz}^*(a^*)$ and hence, the identity
$\partial_z P_{zz}=0$ allows to express $\partial_zp$ in terms of $\partial_zT$. The result is
\beq
\label{6.4}
\frac{\partial \ln p}{\partial z}=-\frac{(1-\beta)a^*(\partial_{a^*} P_{zz}^*)}{P_{zz}^*-a^*(\partial_{a^*} P_{zz}^*)}
\frac{\partial \ln T}{\partial z}.
\eeq
Finally, the balance equation \eqref{3.5} for the granular temperature yields
\beq
\label{6.4.1}
a P_{xy}^{(0)}=-\frac{d}{2}p \zeta^{(0)}.
\eeq
Upon deriving \eqref{6.4.1} we have neglected the term $\partial_z q_z$ since it is of second order in the gradients of $x_1$, $p$ and $T$. As said in section \ref{sec2}, Eq.\ \eqref{6.4.1} establishes a relation between the (reduced) shear rate $a^*$ and the coefficient of restitution $\alpha_{22}$.

Use of Eq.\ \eqref{6.4} into Eq.\ \eqref{6.3} and substitution of Eq.\ \eqref{6.3} into Eq.\ \eqref{6.1} finally leads to
\beq
\label{6.5}
\Lambda_z=\frac{\overline{D}_{T,zz}-(1-\beta)a^*\left(P_{zz}^*-a^*(\partial_{a^*} P_{zz}^*)\right)^{-1}\overline{D}_{p,zz}
(\partial_{a^*} P_{zz}^*)}{D_{zz}^*},
\eeq
where $\overline{D}_{p,zz}\equiv x_1^{-1}D_{p,zz}^*$ and $\overline{D}_{T,zz}\equiv x_1^{-1}D_{T,zz}^*$. Equation \eqref{6.5} provides the thermal diffusion factor $\Lambda_z$ in terms of the diffusion coefficients $D_{zz}^*$, $D_{p,zz}^*$ and $D_{T,zz}^*$, the (reduced) pressure tensor $P_{zz}^*$ and the derivative $\partial_{a^*} P_{zz}^*$. To evaluate those quantities, we consider model A ($\beta=0$) where $P_{zz}^*$ and $\partial_{a^*} P_{zz}^*$ are given by Eqs.\ \eqref{c1} and \eqref{d1}, respectively.
In addition, the explicit forms of the diffusion coefficients can be found by solving the set of algebraic equations \eqref{5.1}--\eqref{5.3} for $i=j=z$.
The results clearly show that, while $D_{zz}^*>0$, the coefficients $D_{p,zz}^*$ and $D_{T,zz}^*$ do not have a definite sign.

The condition $\Lambda_z=0$ provides the segregation criterion for the upwards/downwards segregation transition. Thus, according to Eq.\ \eqref{6.5} and given that $D_{zz}^*>0$,
the marginal segregation curve ($\Lambda_z=0$) separating segregation towards the cold wall ($\Lambda_z>0$) from segregation towards the hot wall ($\Lambda_z<0$) is given by the condition
\beq
\label{6.6}
\left(P_{zz}^*-a^*(\partial_{a^*} P_{zz}^*)\right)D_{T,zz}^*=a^*(\partial_{a^*} P_{zz}^*) D_{p,zz}^*.
\eeq
\vicente{Although relation \eqref{6.6} holds for models A and B alike, the form of the phase diagrams for segregation ($\Lambda_z=0$) depends on the interaction parameter $\beta$,
since the quantities $\partial_{a^*} P_{zz}^*$, $D_{T,zz}^*$ and  $D_{p,zz}^*$ differ in both models, even in the steady state. On the other hand, according to the previous
results derived in the monodisperse case \cite{G07}, it is expected that the influence of $\beta$ on segregation is very weak.}

Before analyzing the dependence of the parameter space on the form of the phase diagrams, it is instructive to consider some limit situations.
When the intruder and the particles of the gas are mechanically equivalent ($m_1=m_2$, $\sigma_1=\sigma_2$ and $\alpha_{22}=\alpha_{12}$), the two species do not segregate.
This is consistent with Eq. \ \eqref{6.6} since then $D_{p,zz}^*=D_{T,zz}^*=0$ so that $\Lambda_z=0$ for any value of the (common) coefficient of restitution.
Another interesting situation is the elastic limit ($\alpha_{22}=\alpha_{12}=1$, which implies $a^*=0$ in the steady state condition \eqref{6.4.1}).
In this case, $P_{zz}^*=1$ and $D_{T,zz}^*=0$ so that, Eq.\ \eqref{6.6} holds trivially for any value of the ratios $m_1/m_2$ and $\sigma_1/\sigma_2$ (the intruder does not segregate).
Beyond the above two limiting cases, the criterion \eqref{6.6} is rather complicated since it involves all the parameter space of the problem ($m_1/m_2$, $\sigma_1/\sigma_2$, $\alpha_{22}$).

Figure \ref{fig5} shows the phase diagram in the $\left\{m_1/m_2, \sigma_1/\sigma_2\right\}$ plane for $d=3$ and three different values of the (common) coefficient of restitution $\alpha_{22}=\alpha_{12}$. \vicente{For the sake of simplicity, we consider model A ($\beta=0$).} All zero contours of $\Lambda_z$ pass through the point $(1,1)$ since when $m_1=m_2$ and $\sigma_1=\sigma_2$ all
the species are indistinguishable for this system. We observe that when the intruder is smaller than the gas particles ($\sigma_1<\sigma_2$), the main effect of
collisional dissipation (or equivalently the dimensionless shear rate $a^*$) is to reduce the size of the down segregation region while the opposite happens
when $\sigma_1>\sigma_2$. On the other hand, the impact of dissipation on the latter case is smaller than in the former case (when $\sigma_1<\sigma_2$) and the
curves tend to collapse into a common one for sufficiently large values of the diameter ratio. It is also quite apparent that in general large intruders tend
to move towards colder regions since the upwards segregation is dominant and occupies most of the parameter space. This conclusions contrasts with the results
obtained for vibrated dense systems since intruders tend to move towards hotter regions as they get larger \cite{G08}. It is also important to remark that the
conclusions drawn here for IMM agrees quite well with those obtained before for IHS (see Fig.\ 5 of Ref.\ \cite{GV10}), showing again the reliability of IMM to describe segregation in granular flows.

As a complement of Fig.\ \ref{fig5}, Fig.\ \ref{fig6} shows a phase diagram for $\alpha_{22}\neq \alpha_{12}$ ($\alpha_{22}=0.9$ and $\alpha_{12}=0.7$) \vicente{in the case $\beta=0$}. The theoretical results
derived for IMM are compared here against recent computer simulations performed in Ref.\ \cite{VGK14} in the so-called LTu state, namely, a steady state where the inelastic
cooling is exactly balanced by viscous heating (as in the steady USF state) resulting in a \emph{uniform} heat flux \cite{VSG10,rque51}. In the simulations, segregation is induced
by a thermal gradient parallel to the $y$-direction ($\partial_xT=\partial_zT=0$ but $\partial_y T \neq 0$) so that, the physical situation slightly differs from the one
studied here theoretically. Nevertheless, when $\sigma_1 \approx \sigma_2$ the agreement with theory is good. More significant discrepancies
appear when the intruder is larger than the gas particles since in this case the theory predicts that intruders only move towards hotter regions (upwards segregation).
This contrasts with simulation data since they still show a small region of downwards segregation.

\section{Conclusions}
\label{sec7}

In conclusion, we have investigated the mass, momentum, and heat fluxes for
a binary mixture of inelastic grains. The system is driven out of equilibrium
by an imposed shear flow, which injects energy while dissipative collisions
between the grains act as an energy sink. A kinetic theory description was proposed,
where the intractable Boltzmann equation is simplified in a Maxwell model fashion.
Such models are in some cases simple enough to be amenable to a full analytical
solution, while remaining true to the key physical phenomena under scrutiny.
In this respect, our model is not the simplest possible of the Maxwell family
(the so-called ``plain vanilla'' approach), since the collision frequencies $\omega_{rs}$
are taken to be the same as those found for IHS, see Eq.
(\ref{2.5}). In this equation, a free parameter $\beta$ is introduced. While $\beta=1/2$
is the natural choice to reproduce inelastic hard sphere phenomenology,
it also leads to a complex interplay between shear and dissipation in the steady
state. On the other hand, it is convenient to decouple these effects,
which is possible when $\beta=0$. We thus discriminate two sub-models, referred to
as model A and model B, having respectively $\beta=0$ and $\beta\neq 0$.
Model A enjoys a larger parameter space than model B, which is at the root
of the greater analytical tractability of the approach.

%When restricted to the steady state, the results derived with model A coincide with those of model B.

Perturbing the USF, we analyzed the response of the fluid mixture,
from which generalized transport coefficients can be identified. Due to the anisotropy induced
by the shear, these quantities appear in tensorial, rather than scalar form.
A Chapman-Enskog-like method \vicente{around the shear flow distribution} allows to derive \vicente{the nonlinear differential
equations obeying the set of generalized transport coefficients (see Sec.\ \ref{sec4}). Hopefully, in the case of model A ($\beta=0$), the above equations become simple coupled algebraic equations whose solution} unveil the dependence of transport coefficients on the key parameters (shear, dissipation, concentration, size and mass ratio).
\begin{figure}
\includegraphics[width=0.75 \columnwidth,angle=0]{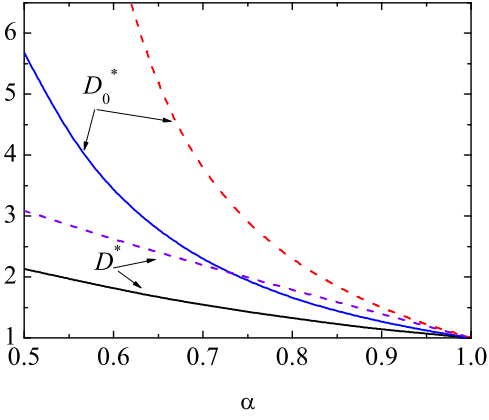}
\caption{(Color online) Plot of the scalar diffusion coefficient $D^*(\al)$ (relative to its elastic value) and the zero shear-rate
diffusion coefficient $D_0^*$ (relative to its elastic value) as functions of the (common) coefficient of restitution $\al_{12}=\al_{22}\equiv \al$ in the steady state for $d=3$ and two different systems: $\sigma_1/\sigma_2=2$ and $m_1/m_2=2$ (solid lines) and $\sigma_1/\sigma_2=1$ and $m_1/m_2=3$ (dashed lines).
\label{fig8}}
\end{figure}

To reduce the complexity of the problem \vicente{and to illustrate the impact of both shearing and collisional dissipation
on transport}, we focussed on the limit where
species 1 has a much smaller concentration than species 2, the so-called
tracer limit ($x_1 \to 0$). In doing so, mass transport becomes the
relevant phenomenon to address, since momentum and heat fluxes coincide
with their mono-component (inelastic) expressions. There are then in general
15 different diffusion transport coefficients, that couple the mass flux to the
gradients of density, pressure and temperature. The simplified model worked out
here reduces this number to 12 (two diagonal and two off-diagonal elements for
each diffusion matrix). Our results hold for arbitrary values of the
shear rate, and are not restricted to small dissipation. They show that
shear driving notably affects mass transport. In addition, good agreement
is reported between our Maxwell treatment and previously derived inelastic hard sphere
results (here, the relevant view is that of model B, where in the steady state,
dissipation selects a unique reduced shear rate).
Finally, we analyzed the segregation phenomenon of an intruder by thermal diffusion
within our framework, which deciphers how shear impinges on the separating effect
of a thermal gradient, opposed by the remixing action of diffusion. Our predictions
are in fair agreement with inelastic hard sphere simulations for Couette flows
sustaining a uniform heat flux.

\vicente{As pointed out in the Introduction, most of the works on granular mixtures \cite{NSIHS,GA05} have been derived by taking the so-called homogeneous cooling
state as the reference state. In this case, the mass transport is characterized by the single scalar coefficients $D$, $D_p$ and $D_T$ (see Eqs.\ \eqref{4.10}--\eqref{4.12} for IMM)
instead of the tensorial quantities $D_{ij}$, $D_{p,ij}$ and $D_{T,ij}$ when the system is sheared. Although these scalar coefficients cannot be directly compared with the diffusion tensors obtained here,
it would be interesting to gauge the effect of dissipation on diffusion in both situations (driven sheared case and freely cooling condition).
In Fig.\ \ref{fig8}, we plot the scalar diffusion coefficient $D^*\equiv \frac{1}{3}(D_{xx}^*+D_{yy}^*+D_{zz}^*)$ (which can be understood as a
generalized diffusion coefficient in a sheared mixture) and the zero shear-rate diffusion coefficient $D_0^*$ (defined in Eq.\ \eqref{5.6}) as functions
of the (common) coefficient of restitution in the steady state (where the results for these coefficients apply for models A and B) for $x_1\to 0$ and $d=3$.
We have scaled both coefficients with respect to their elastic values. Given that the reference states in both descriptions (shear flow state against
homogeneous cooling state) are quite different, there are significant quantitative differences between $D^*$ and $D_0^*$. On the other hand, the dependence
of both coefficients on dissipation is qualitatively similar since they increase as $\al$ decreases. This tendency is more important in the freely cooling
case than in the sheared state, in agreement with the results obtained for IHS (see Fig.\ 5 of Ref.\ \cite{G02}).}

\vicente{Finally, we wish to remark that}, at the expense of a further simplification of the Maxwell model (addressing thus the
aforementioned plain vanilla treatment \cite{GT11}), it is of interest to study the impact on transport
of a recently evidenced transition taking place in the intruder limit,
where the minority species rather unexpectedly carries a finite fraction
of the total system's energy. Work along these lines is underway.

\acknowledgments

V. G. acknowledges support of the Spanish
Government through Grant No. FIS2013-42840-P and
of the Junta de Extremadura (Spain) through Grant
No. GR15104, both partially financed by FEDER funds.
V. G. and E. T. also acknowledge funding by the Investissement d'Avenir LabEx PALM program
(grant number ANR-10-LABX-0039-PALM).

\appendix
\section{Chapman--Enskog-like expansion
\label{appA}}

In this Appendix, some technical details on the determination of the first-order
approximation $f_1^{(1)}$ by means of the Chapman--Enskog-like expansion are provided.
Inserting the expansions (\ref{3.8})--(\ref{3.11}) into Eq.\ (\ref{3.1}), one gets the
kinetic equation for $f_1^{(1)}$:
\beqa
\label{a1}
& & \partial_t^{(0)}f_1^{(1)}-a V_y\frac{\partial f_1^{(1)}}{\partial V_x}+{\cal
L}_1 f_1^{(1)}+{\cal M}_1 f_2^{(1)}=
\nonumber\\
& &
-\left[\partial_t^{(1)}+({\bf V}+{\bf u}_0)\cdot
\nabla \right]f_1^{(0)}.
\eeqa
The velocity dependence on the right-hand side of Eq.\ (\ref{a1}) can be obtained from
the macroscopic balance equations (\ref{3.3})--(\ref{3.5}) to first order in the
gradients. Using these balance equations in Eq.\ (\ref{a1}), one gets
\beqa
\label{a7}
& &  \partial_t^{(0)}f_1^{(1)}-a V_y\frac{\partial f_1^{(1)}}{\partial V_x}+{\cal
L}_1 f_1^{(1)}+{\cal M}_1 f_2^{(1)} =
\nonumber\\
& & {\bf A}_1\cdot \nabla x_1
+{\bf B}_1\cdot \nabla p+{\bf C}_1\cdot \nabla T+{\sf D}_1:\nabla \delta {\bf u},
\eeqa
where
\begin{equation}
\label{a8} A_{1,i}({\bf c})=-\frac{\partial f_1^{(0)}}{\partial x_1}c_i-\frac{1}{\rho}
\frac{\partial f_1^{(0)}}{\partial c_j}\frac{\partial P_{ij}^{(0)}}{\partial
x_1},
\end{equation}
\begin{equation}
\label{a9} B_{1,i}({\bf c})=-\frac{\partial f_1^{(0)}}{\partial p} c_i-\frac{1}{\rho}
\frac{\partial f_1^{(0)}}{\partial c_j}\frac{\partial P_{ij}^{(0)}}{\partial p},
\end{equation}
\begin{equation}
\label{a10} C_{1,i}({\bf c})=-\frac{\partial f_1^{(0)}}{\partial T} c_i-\frac{1}{\rho}
\frac{\partial f_1^{(0)}}{\partial c_j}\frac{\partial P_{ij}^{(0)}}{\partial T},
\end{equation}
\beqa
\label{a11}
D_{1,ij}({\bf c})&=&p\frac{\partial f_1^{(0)}}{\partial
p}\delta_{ij}+c_j\frac{\partial f_1^{(0)}} {\partial c_i}+\frac{2}{d
p}\left(P_{ij}^{(0)}-a\eta_{xyij}\right) \nonumber\\
& & \times\left(p\frac{\partial f_1^{(0)}}{\partial
p}+T\frac{\partial f_1^{(0)}}{\partial T}\right).
\eeqa
Upon writing Eq.\ (\ref{a11}) use has been made of the identity $\zeta^{(1)}=0$ and the
expression of the total pressure tensor $P_{ij}^{(1)}$ of the mixture
\begin{equation}
\label{a12} P_{ij}^{(1)}=-\eta_{ijk\ell} \frac{\partial \delta u_k} {\partial
r_{\ell}},
\end{equation}
where $\eta_{ijk\ell}$ is the viscosity tensor.

The solution to Eq.\ (\ref{a7}) has the form given by Eq.\ (\ref{3.17}), where the
coefficients ${\boldsymbol {\cal A}}_{1}$, ${\boldsymbol {\cal B}}_{1}$, ${\boldsymbol
{\cal C}}_{1}$, and ${\sf {\cal D}}_{1}$ are functions of the peculiar velocity and the
hydrodynamic fields $x_1$, $p$, $T$, and $\delta {\bf u}$. The time derivative acting
on these quantities can be evaluated with the replacement
\begin{equation}
\label{a14} \partial_t^{(0)}\to -\left(\frac{2}{d p}a
P_{xy}^{(0)}+\zeta^{(0)}\right)\left(p\partial_p+T\partial_T\right).
\end{equation}
Moreover, there are contributions from $\partial_t^{(0)}$ acting on the pressure,
temperature, and velocity gradients given by
\begin{eqnarray}
\label{a15}
\partial_t^{(0)} \nabla p&=&-\nabla \left(\frac{2}{d}a
P_{xy}^{(0)}+p\zeta^{(0)}\right)
\nonumber\\
&=&-\left(\frac{2a}{d}\frac{\partial P_{xy}^{(0)}}{\partial x_1}+p\frac{\partial
\zeta^{(0)}}{\partial x_1}\right)\nabla x_1\nonumber\\
& & - \left(\frac{2a}{d} \frac{\partial
P_{xy}^{(0)}}{\partial p}+ \zeta^{(0)}+\frac{\partial
\zeta^{(0)}}{\partial p}\right) \nabla p \nonumber\\
& &-\left(\frac{2a}{d} \frac{\partial P_{xy}^{(0)}}{\partial T}
+p\frac{\partial
\zeta^{(0)}}{\partial T}\right) \nabla T,
\end{eqnarray}
\begin{eqnarray}
\label{a16}
\partial_t^{(0)} \nabla T&=&-\nabla \left(\frac{2T}{d p}a
P_{xy}^{(0)}+T\zeta^{(0)}\right)
\nonumber\\
&=&-\left(\frac{2aT}{d p}\frac{\partial P_{xy}^{(0)}}{\partial x_1}+T\frac{\partial
\zeta^{(0)}}{\partial x_1}\right)\nabla x_1\nonumber\\
&+& \left(\frac{2aT}{d p^2}P_{xy}^{(0)}-
\frac{2aT}{d p}\frac{\partial P_{xy}^{(0)}}{\partial p}- T\frac{\partial
\zeta^{(0)}}{\partial p}\right) \nabla p \nonumber\\
&-&\left(\frac{2a}{d p}P_{xy}^{(0)}+\frac{2aT}{d p} \frac{\partial
P_{xy}^{(0)}}{\partial T}+\zeta^{(0)}+T\frac{\partial \zeta^{(0)}}{\partial T}\right)
\nabla T,
\nonumber\\
\end{eqnarray}
\begin{equation}
\label{a17}
\partial_t^{(0)} \nabla_i \delta u_j=\nabla_i \partial_t^{(0)} \delta u_j=-a_{jk} \nabla_i \delta u_k.
\end{equation}
The corresponding integral equations (\ref{3.18})--(\ref{3.20}) can be obtained when
one identifies coefficients of independent gradients in Eq.\ (\ref{a7}) and takes into
account Eqs.\ (\ref{a15})--(\ref{a17}) and the mathematical property
\begin{eqnarray}
\label{a18}
\partial_t^{(0)} X &=&\frac{\partial X}{\partial p}\partial_t^{(0)}
p+\frac{\partial X}{\partial T}\partial_t^{(0)} T+\frac{\partial X}{\partial \delta
u_i}\partial_t^{(0)}
\delta u_i \nonumber\\
&=&-\left(\frac{2}{d p}a P_{xy}^{(0)}+\zeta^{(0)}\right)\left(p\frac{\partial X}{\partial p}+T\frac{\partial X}
{\partial T}\right)\nonumber\\
& & +a_{ij}\delta u_j \frac{\partial X}{\partial c_i},
\end{eqnarray}
where in the last step it has been taken into account that $X$ depends on $\delta {\bf
u}$ through ${\bf c}={\bf V}-\delta {\bf u}$.

\section{Heat flux transport coefficients}
\label{appB}

The heat flux is defined by Eq.\ (\ref{3.26}) in terms of
the coefficients $D_{ij}''$ (Eq.\ (\ref{3.31})), $L_{ij}$ (Eq.\ (\ref{3.32})) and
$\lambda_{ij}$ (Eq.\ (\ref{3.33})). In order to determine them, we introduce the quantities
\begin{equation}
\label{b1}
D_{r,ijk\ell}''=-\frac{m_r}{2T^2}\int\dd{\bf c}\; c_ic_jc_k {\cal A}_{r,\ell}({\bf c}),
\end{equation}
\begin{equation}
\label{b2}
L_{r,ijk\ell}=-\frac{m_r}{2}\int\dd{\bf c}\; c_ic_jc_k {\cal B}_{r,\ell}({\bf c}),
\end{equation}
\begin{equation}
\label{b3}
\lambda_{r,ijk\ell}=-\frac{m_r}{2}\int\dd{\bf c}\; c_ic_jc_k {\cal C}_{r,\ell}({\bf c}).
\end{equation}
The generalized transport coefficients $D_{ij}''$, $L_{ij}$ and $\lambda_{ij}$ are
defined as
\begin{equation}
\label{b4} D_{ij}''=\sum_{s=1}^2\; D_{s,kkij}'',\quad L_{ij}=\sum_{s=1}^2\;
L_{s,kkij}, \quad \lambda_{ij}=\sum_{s=1}^2\; \lambda_{s,kkij}.
\end{equation}

The differential equations verifying the (scaled) coefficients $D_{r,ijk\ell}^*\equiv D_{r,ijk\ell}''(m_1+m_2)\nu_0/n$, $L_{r,ijk\ell}^*=L_{r,ijk\ell}(m_1+m_2)\nu_0/T$ and
$\lambda_{r,ijk\ell}^*=\lambda_{r,ijk\ell}(m_1+m_2)\nu_0/p$ can be obtained by following similar mathematical steps as those made for the other transport coefficients.
The final results can be written as
\begin{widetext}
\begin{eqnarray}
\label{b5}& & \left(\frac{2a^*}{d} P_{xy}^{*}+\zeta^{*}\right)
\left[2-\beta(1+a^*\partial_{a^*}\right]D_{1,ijk\ell}^*-A D_{1,ijk\ell}^*-B
\left[D_{2,ijk\ell}^*-\frac{1}{2}\left(\delta_{kj}D_{2,i\ell}^*+\delta_{ij}D_{2,k\ell}^*+\delta_{ik}D_{2,j\ell}^*\right)
\right]\nonumber\\
& &
-C\left(\delta_{kj}D_{1,i\ell}^*+\delta_{ij}D_{1,k\ell}^*+\delta_{ik}D_{1,j\ell}^*\right)
-E\left(\delta_{kj}D_{i\ell}^*+\delta_{ij}D_{k\ell}^*+\delta_{ik}D_{j\ell}^*\right)\nonumber\\
& &
-a^*\left(\delta_{ix}D_{1,jky\ell}^*+\delta_{jx}D_{1,iky\ell}^*+\delta_{kx}D_{1,ijy\ell}^*\right)=
-\left(L_{1,ijk\ell}^*+\lambda_{1,ijk\ell}^*\right)\left(\frac{2a^*}{d}\partial_{x_1} P_{xy}^{*}+\partial_{x_1}\zeta^*\right)\nonumber\\
& & -\frac{1}{T^2\nu_0D_0''}\partial_{x_1}N_{1,ijk\ell}^{(0)}+\frac{n(m_1+m_2)}{\rho}
\left( P_{1,kj}^*\partial_{x_1}P_{i\ell}^*+P_{1,ik}^*\partial_{x_1}P_{j\ell}^*
+P_{1,ij}^*\partial_{x_1}P_{k\ell}^*\right),
\end{eqnarray}
\begin{eqnarray}
\label{b6}& & \left(\frac{2a^*}{d} P_{xy}^{*}+\zeta^{*}\right)
\left[1-\beta(1+a^*\partial_{a^*})\right]L_{1,ijk\ell}^*-A L_{1,ijk\ell}^*-B \left[
L_{2,ijk\ell}^*-\frac{1}{2}\left(\delta_{kj}L_{2,i\ell}^*+\delta_{ij}L_{2,k\ell}^*+\delta_{ik}L_{2,j\ell}^*\right)
\right]\nonumber\\
& &
-C \left(\delta_{kj}L_{1,i\ell}^*+\delta_{ij}L_{1,k\ell}^*+\delta_{ik}L_{1,j\ell}^*\right)
-E \left(\delta_{kj}D_{p,i\ell}^*+\delta_{ij}D_{p,k\ell}^*+\delta_{ik}D_{p,j\ell}^*\right)\nonumber\\
& &
-a^*\left(\delta_{ix}L_{1,jky\ell}^*+\delta_{jx}L_{1,iky\ell}^*+\delta_{kx}L_{1,ijy\ell}^*\right)+
L_{1,ijk\ell}^*\left[\frac{2a^*}{d}(1-a^*\partial_{a^*})P_{xy}^{*}+ (2-a^*\partial_{a^*}) \zeta^{*}
\right]\nonumber\\
& =& -\frac{1}{\nu_0L_0}\partial_{p}N_{1,ijk\ell}^{(0)}+\frac{n(m_1+m_2)}{2\rho}\left[
P_{1,kj}^*(1-a^*\partial_{a^*})P_{i\ell}^*+P_{1,ik}^*(1-a^*\partial_{a^*})P_{j\ell}^*
+P_{1,ij}^*(1-a^*\partial_{a^*})P_{k\ell}^*\right]\nonumber\\
& &+\lambda_{1,ijk\ell}^*\left[\frac{2a^{*2}}{d}\partial_{a^*}P_{xy}^{*}- (1-a^*\partial_{a^*})\zeta^{*}
\right],
\end{eqnarray}
\begin{eqnarray}
\label{b7}& & \left(\frac{2a^*}{d} P_{xy}^{*}+\zeta^{*}\right)
\left[1-\beta(1+a^*\partial_{a^*})\right]\lambda_{1,ijk\ell}^*-A
\lambda_{1,ijk\ell}^*-B \left[
\lambda_{2,ijk\ell}^*-\frac{1}{2} \left(\delta_{kj}\lambda_{2,i\ell}^*+\delta_{ij}\lambda_{2,k\ell}^*+\delta_{ik}
\lambda_{2,j\ell}^*\right)\right]\nonumber\\
& &
-C \left(\delta_{kj}\lambda_{1,i\ell}^*+\delta_{ij}\lambda_{1,k\ell}^*+
\delta_{ik}\lambda_{1,j\ell}^*\right)
-E \left(\delta_{kj}D_{T,i\ell}^*+\delta_{ij}D_{T,k\ell}^*+\delta_{ik}D_{T,j\ell}^*
\right)\nonumber\\
& &
-a^*\left(\delta_{ix}\lambda_{1,jky\ell}^*+\delta_{jx}\lambda_{1,iky\ell}^*+
\delta_{kx}\lambda_{1,ijy\ell}^*\right)+
\lambda_{1,ijk\ell}^*\left[\frac{2a^*}{d}\left(1+(1-\beta)a^*\partial_a^*
\right)P_{xy}^{*}+ \left(\beta+ (1-\beta)a^*\partial_{a^*}\right)\zeta^*\right]\nonumber\\
& =&
-\frac{1}{\nu_0\lambda_0}\partial_{T}N_{1,ijk\ell}^{(0)}+\frac{n(m_1+m_2)}{2\rho}
(1-\beta)a^*\left[
P_{1,kj}^*\partial_{a^*}P_{i\ell}^*+P_{1,ik}^*\partial_{a^*}P_{j\ell}^*
+P_{1,ij}^*\partial_{a^*}P_{k\ell}^*\right]\nonumber\\
& &-L_{1,ijk\ell}^*\left[\frac{2a^{*2}}{d}(1-\beta) \partial_a^*P_{xy}^{*}+(\beta-1)(1-a^*\partial_a^*)
\zeta^{*}\right].
\end{eqnarray}
\end{widetext}
In Eqs.\ \eqref{b5}--\eqref{b7}, we have introduced the fourth-degree velocity
moments of the zeroth-order distribution $f_r^{(0)}$,
\begin{equation}
\label{b8} N_{r,ijk\ell}^{(0)}=\frac{m_r}{2}\int\dd{\bf c}\; c_ic_jc_k c_\ell
f_r^{(0)}({\bf c}),
\end{equation}
and the (dimensionless) quantities
\beqa
\label{b9}
A &=&\frac{3}{2}\omega_{11}^*(1+\alpha_{11})\frac{1+d-\alpha_{11}}{d(d+2)}+
\frac{3}{d}\omega_{12}^*\mu_{21}(1+\alpha_{12})\nonumber\\
& & \times\left[1-\frac{2\mu_{21}(1+\alpha_{12})}
{d+2}\left(1-\frac{\mu_{21}(1+\alpha_{12})}{d+4}\right)\right],
\nonumber\\
\eeqa
\begin{equation}
\label{b10}
B =-6\frac{\rho_1}{\rho_2}\omega_{12}^*\frac{\mu_{21}^3(1+\alpha_{12})^3}{d(d+2)(d+4)},
\end{equation}
\beqa
\label{b11}
C &=&-\frac{13+4d-3\alpha_{11}}{8d(d+2)(d+4)}\omega_{11}^*(1+\alpha_{11})^2
+2\frac{\omega_{12}^*}{d(d+2)(d+4)}
\nonumber\\
& & \times \mu_{21}^2(1+\alpha_{12})^2\left[3\mu_{21}(1+\alpha_{12})-(d+4) \right],
\eeqa
\begin{eqnarray}
\label{b12} E&=&\omega_{11}^*\frac{(1-\alpha_{11}^2)}{8d(d+2)}\frac{\gamma_1}{\mu_{12}}\left(
3\alpha_{11}-d+1\right)-\frac{\omega_{12}^*}{2d(d+2)}\nonumber\\
& & \times \frac{x_1\gamma_1}{x_2
\mu_{21}}\left[d+2+3\mu_{21}(1+\alpha_{12})
\left(\mu_{21}(1+\alpha_{12})-2\right)\right. \nonumber\\
& & \left.
+\frac{x_2\gamma_2}{x_1\gamma_1}\mu_{21}(1+\alpha_{12})\left(3\mu_{21}(1+\alpha_{12})
-d-2\right)\right].
\nonumber\\
\end{eqnarray}
In Eqs.\ \eqref{b9}--\eqref{b12}, $\omega_{rs}^*\equiv \omega_{rs}/\nu_0$. The differential equations for the coefficients $D_{2,ijk\ell}^*$, $L_{2,ijk\ell}^*$ and $\lambda_{2,ijk\ell}$ can be
 obtained from Eqs.\ \eqref{b5}--\eqref{b7} by changing $1\leftrightarrow 2$. As in the case of the previous transport coefficients, Eqs.\ \eqref{b5}--\eqref{b7} become algebraic for model A ($\beta=0$).
Even for this model, the solution to the above equations requires the knowledge of the fourth degree moments $N_{r,ijk\ell}^{(0)}$ whose expressions are only known for a monodisperse granular gas of
IMM \cite{SG07}.

\section{Rheological properties in the USF. Tracer limit}
\label{appC}

The explicit forms of the (reduced) pressure tensors $P_{2,ij}^*\equiv P_{2,ij}/n_2T_2$ and $P_{1,ij}^*\equiv P_{1,ij}/n_2T_2$ of the solvent (excess) and the solute (tracer) components, respectively,
of a granular binary mixture (in the tracer limit $x_1\to 0$) of IMM under USF are provided in this Appendix. We consider here model A ($\beta=0$) where the coefficients of restitution $\alpha_{rs}$ and
the (reduced) shear rate $a^*$ are decoupled.

The non-zero elements of $P_{2,ij}^*$ are given by \cite{G07}
\beq
\label{c1}
P_{2,yy}^*=P_{2,zz}^*=\frac{1}{1+2\Lambda(\widetilde{a})}, \quad
P_{2,xx}^*=\frac{1+2d \Lambda(\widetilde{a})}{1+2\Lambda(\widetilde{a})},
\eeq
\beq
\label{c2}
P_{2,xy}^*=-\frac{\widetilde{a}}{\left[1+2\Lambda(\widetilde{a})\right]^2},
\eeq
where
\beq
\label{c3}
\widetilde{a}=\frac{2(d+2)}{(1+\al_{22})^2}\frac{a^*}{\omega_{22}^*},
\eeq
\beq
\label{c5.1}
\omega_{22}^*\equiv \frac{\omega_{22}}{\nu_0}=\sqrt{2\mu_{12}}\left(\frac{\sigma_2}{\sigma_{12}}\right)^{d-1},
\eeq
and $\Lambda(\widetilde{a})$ is the real root of the cubic equation
\beq
\label{c4}
\Lambda(1+2\Lambda)^2=\frac{\widetilde{a}^2}{d},
\eeq
namely
\beq
\label{c5}
\Lambda(\widetilde{a})=\frac{2}{3}\sinh^2\left[\frac{1}{6}\cosh^{-1}\left(1+\frac{27}{d}
\widetilde{a}^2\right)\right].
\eeq
In addition, the long-time behavior of the granular temperature $T(t)\simeq T_2(t)$ is $T_2(t)=T_2(0)e^{\lambda \omega_{22} t}$ where
\beq
\label{c6}
\lambda =-\frac{2}{d}\frac{P_{2,xy}^*a^*}{\omega_{22}^*}-\frac{\zeta^{*}}{\omega_{22}^*}=
\frac{(1+\al_{22})^2}{d+2}\Lambda-
\frac{1-\al_{22}^2}{2d\omega_{22}^*}.
\eeq
Upon obtaining the second identity in \eqref{c6} use has been made of Eq.\ \eqref{c2} and the result $\zeta^*=(1-\al_{22}^2)/2d$.

In the case of tracer particles, the relevant elements of $P_{1,ij}^*$ can be written as \cite{GT10}
\beq
\label{c7}
P_{1,yy}^*=P_{1,zz}^*=x_1 \frac{F+H P_{2,yy}^*}{\lambda \omega_{22}^*+G},
\eeq
\beq
\label{c8}
P_{1,xy}^*=x_1 \frac{H P_{2,xy}^*-a^*x_1^{-1} P_{1,yy}^*}{\lambda \omega_{22}^*+G},
\eeq
\beq
\label{c8.1}
P_{1,xx}^*= x_1\frac{F+H P_{2,xx}^*-2a^* x_1^{-1}P_{1,xy}^*}{\lambda \omega_{22}^*+G},
\eeq
where
\beq
\label{c9}
F=\frac{\omega_{12}^*}{d+2}\mu_{21}\left(\mu_{12}+\mu_{21}\gamma\right)(1+\al_{12})^2,
\eeq
\beq
\label{c10}
G=\frac{2\omega_{12}^*}{d(d+2)}\mu_{21}(1+\al_{12})\left[d+2-\mu_{21}(1+\al_{12})\right],
\eeq
\beq
\label{c11}
H=\frac{2\omega_{12}^*}{d(d+2)}\mu_{21}\mu_{12}(1+\al_{12})^2.
\eeq
Here,
\beq
\label{c11.1}
\omega_{12}^*\equiv \frac{\omega_{12}}{\nu_0}=\sqrt{\mu_{12}+\mu_{21}\gamma},
\eeq
where $\gamma \equiv T_1/T_2$ is the temperature ratio. The temperature ratio is determined from the constraint
\beq
\label{c12}
x_1\gamma=\frac{P_{1,xx}^*+(d-1)P_{1,yy}^*}{d}.
\eeq
Since the collision frequency $\omega_{12}^*$ is a nonlinear function of $\gamma$,
one then has to numerically solve Eq.\ \eqref{c12} to obtain the shear-rate dependence of the temperature ratio.

\section{Evaluation of the derivatives of the pressure tensors with respect to the shear rate. Tracer limit}
\label{appD}

This Appendix addresses the evaluation of the derivatives $\partial_{a^*} P_{2,ij}^*$ and $\partial_{a^*} P_{1,ij}^*$ for model A ($\beta=0$) needed to determine the tracer diffusion coefficients
$D_{ij}$, $D_{p,ij}$ and $D_{T,ij}$ in the tracer limit. In the case of the excess component, according to Eqs.\ \eqref{c1} and \eqref{c2}, one has \cite{G07}
\beq
\label{d1}
a^* \frac{\partial P_{2,yy}^*}{\partial a^*}=-\frac{4\Lambda}{(1+2\Lambda)(1+6\Lambda)},
\eeq
\beq
\label{d1.1}
a^* \frac{\partial P_{2,xx}^*}{\partial a^*}=\frac{4(d-1)\Lambda}{(1+2\Lambda)(1+6\Lambda)},
\eeq
\beq
\label{d2}
a^* \frac{\partial P_{2,xy}^*}{\partial a^*}=-\frac{1-2\Lambda}{(1+2\Lambda)^2(1+6\Lambda)}\widetilde{a},
\eeq
where use has been made of the identity
\beq
\label{d3}
a^* \frac{\partial \Lambda}{\partial a^*}=2\Lambda \frac{1+2\Lambda}{1+6\Lambda}.
\eeq

The calculations for the tracer particles are more intricate. First, we derive both sides of Eq.\ \eqref{c7} with respect to $a^*$ to obtain the result
\beq
\label{d4}
\frac{\partial P_{1,yy}^*}{\partial a^*}=\Delta_{yy}^{(0)}+\Delta_{yy}^{(1)}\frac{\partial \gamma}{\partial a^*},
\eeq
where
\beq
\label{d4.1}
\Delta_{yy}^{(0)}=\frac{x_1H(\partial_{a^*}P_{2,yy}^*)-P_{1,yy}^*\omega_{22}^*(\partial_{a^*}\lambda)}
{\lambda \omega_{22}^*+G},
\eeq
\beq
\label{d4.2}
\Delta_{yy}^{(1)}=\frac{x_1}{\lambda \omega_{22}^*+G}\left(\overline{F}+\overline{H}P_{2,yy}^*
-\overline{G}x_1^{-1}P_{1,yy}^*\right).
\eeq
In Eqs.\ \eqref{d4.1} and \eqref{d4.2}, $\partial_{a^*}\lambda=\frac{1+\al^2}{d+2}(\partial_{a^*}\Lambda)$ and
we have introduced the quantities
\beq
\label{d5}
\overline{F}\equiv \frac{\mu_{21}}{2\omega_{12}^{*2}}F+\omega_{12}^*\frac{\mu_{21}^2(1+\al_{12})^2}{d+2},
\eeq
\beq
\label{d6}
\overline{G}\equiv \frac{\mu_{21}}{2\omega_{12}^{*2}}G, \quad \overline{H}\equiv \frac{\mu_{21}}{2\omega_{12}^{*2}}H.
\eeq
The derivatives $\partial_{a^*} P_{1,xy}^*$ and $\partial_{a^*} P_{1,xx}^*$ can be also obtained from Eqs.\ \eqref{c8} and \eqref{c8.1}. Their final forms can be written as
\beq
\label{d7}
\frac{\partial P_{1,xy}^*}{\partial a^*}=\Delta_{xy}^{(0)}+\Delta_{xy}^{(1)}\frac{\partial \gamma}{\partial a^*},
\eeq
\beq
\label{d8}
\frac{\partial P_{1,xx}^*}{\partial a^*}=\Delta_{xx}^{(0)}+\Delta_{xx}^{(1)}\frac{\partial \gamma}{\partial a^*},
\eeq
where
\begin{widetext}
\beq
\label{d9}
\Delta_{xy}^{(0)}=\frac{x_1H(\partial_{a^*}P_{2,xy}^*)-P_{1,yy}^*-a^* \Delta_{yy}^{(0)}-P_{1,xy}^*\omega_{22}^*(\partial_{a^*}\lambda)}
{\lambda \omega_{22}^*+G},
\eeq
\beq
\label{d10}
\Delta_{xy}^{(1)}=\frac{x_1}{\lambda \omega_{22}^*+G}\left(\overline{H}P_{2,xy}^*-a^* x_1^{-1}\Delta_{yy}^{(1)}
-\overline{G}x_1^{-1}P_{1,xy}^*\right),
\eeq
\beq
\label{d11}
\Delta_{xx}^{(0)}=\frac{x_1H(\partial_{a^*}P_{2,xx}^*)-2P_{1,xy}^*-2a^* \Delta_{xy}^{(0}-P_{1,xx}^*\omega_{22}^*(\partial_{a^*}\lambda)}
{\lambda \omega_{22}^*+G},
\eeq
\beq
\label{d12}
\Delta_{xx}^{(1)}=\frac{x_1}{\lambda \omega_{22}^*+G}\left(\overline{F}+\overline{H}P_{2,xx}^*-2a^* x_1^{-1}\Delta_{xy}^{(1)}
-\overline{G}x_1^{-1}P_{1,xx}^*\right).
\eeq
\end{widetext}

To close the problem, it still remains to get the quantity $\partial_{a^*}\gamma$, which can be determined from the relation \eqref{c12} by taking the derivative with respect to $a^*$
in both sides of this identity. The result can be written as  \beq
\label{d13}
\frac{\partial \gamma}{\partial a^*}=x_1^{-1}\frac{\Delta_{xx}^{(0)}+(d-1)\Delta_{yy}^{(0)}}{d-x_1^{-1}\Delta_{xx}^{(1)}
-(d-1)x_1^{-1}\Delta_{yy}^{(1)}}.
\eeq


\begin{thebibliography} {99}
\bibitem{rque51}
\vicente{Thermal diffusion is induced by a thermal gradient; in the USF state, temperature is uniform and one cannot determine the thermal diffusion factor.
We thus consider a state where there is a coexistence between a (linear) shear field and a temperature gradient.
With the resulting uniform heat flux, the steady condition (\ref{6.4.1}) still applies.}
\end{thebibliography}
\end{document}